\documentclass[12pt,a4paper]{article}
\usepackage{graphicx}
\usepackage[T1]{fontenc}
\usepackage[utf8]{inputenc}
\usepackage{textcomp}
\usepackage[sc,osf]{mathpazo}
\usepackage{a4wide}  
\usepackage{latexsym,amsthm,amsfonts,amsmath,mathrsfs,amssymb}
\usepackage{dsfont}
\usepackage{accents}
\usepackage[nosort]{cite}
\usepackage{booktabs} 
\usepackage[unicode,implicit]{hyperref}
\hypersetup{%
  pdftitle    = {Generalized Komar charges and Smarr formulas
    for black holes and boson stars}
  pdfkeywords = {Black holes, scalar fields, scalar charges, hair, no-hair
    theorems, symmetry, charge, Smarr formula},
  pdfauthor   = {Romina Ballesteros and Tom\'as Ort\'{\i}n},
  plainpages  = true,
  colorlinks  = true,
  citecolor   = blue,
  urlcolor    = red,
  linkcolor   = black
}
\newcommand{\hepth}[1]{{\tt
\href{http://www.arXiv.org/abs/hep-th/#1}{hep-th/#1}}}
\newcommand{\grqc}[1]{{\tt
\href{http://www.arXiv.org/abs/gr-qc/#1}{gr-qc/#1}}}

\newcommand{\arxiv}[1]{{\tt arXiv:\href{http://www.arXiv.org/abs/#1}{#1}}}
\newcommand{\doi}[1]{{\tt DOI:\href{http://dx.doi.org/#1}{#1}}}

\allowdisplaybreaks

\makeatletter
\@addtoreset{equation}{section}
\makeatother

\begin{document}

\begin{flushright}
\small
IFT-UAM/CSIC-24-128\\
September 12\textsuperscript{th}, 2024\\
Revised: February 28\textsuperscript{th}, 2025
\normalsize
\end{flushright}

\vspace{1cm}

\begin{center}

  {\Large {\bf Generalized Komar charges and Smarr formulas\\[.5cm]
    for black holes and boson stars}}

\vspace{1.5cm}

\renewcommand{\thefootnote}{\alph{footnote}}

{\sl\large Romina Ballesteros$^{1,2,}$}\footnote{Email: {\tt romina.ballesteros[at]estudiante.uam.es}}
{\sl\large and Tom\'as Ort\'{\i}n}$^{1,}$\footnote{Email: {\tt  tomas.ortin[at]csic.es}}

\setcounter{footnote}{0}
\renewcommand{\thefootnote}{\arabic{footnote}}
\vspace{1cm}

${}^{1}${\it Instituto de F\'{\i}sica Te\'orica UAM/CSIC\\
C/ Nicol\'as Cabrera, 13--15,  C.U.~Cantoblanco, E-28049 Madrid, Spain}\\

\vspace{.5cm}

$^{2}${\it Pontificia Universidad Cat\'olica de Valpara\'{\i}so,
  Instituto de F\'{\i}sica, Av. Brasil 2950, Valpara\'{\i}so, Chile}

\vspace{1cm}


{\bf Abstract}
\end{center}
\begin{quotation}
  {\small The standard Komar charge is a $(d-2)$-form that can be defined in
    spacetimes admitting a Killing vector and which is closed when the vacuum
    Einstein equations are satisfied. Its integral at spatial infinity (the
    Komar integral) gives the conserved charge associated to the Killing
    vector, and, due to its on-shell closedness, the same value (expressed in
    terms of other physical variables) is obtained integrating over the event
    horizon (if any). This equality is the basis of the Smarr formula.  This
    charge can be generalized so that it still is closed on-shell in presence
    of matter and its integrals give generalizations of the Smarr formula.  We
    show how the Komar charge and other closed $(d-2)$-form charges can be
    used to prove non-existence theorems for gravitational solitons and boson
    stars. In particular, we show how one can deal with generalized symmetric
    fields (invariant under a combination of isometries and other global
    symmetries) and how the generalized symmetric ansatz permits to evade the
    non-existence theorems.  }
\end{quotation}

\newpage
\pagestyle{plain}




\section{Introduction}

Charge conservation is one of the more powerful ideas of Physics. Conserved
charges can be used to label and characterize states or whole physical systems
and their presence constrains their evolution and simplifies the study of
their dynamics.

The conservation of many charges is associated, via Noether's theorem, to
global symmetries. For each of those global symmetries there is a conserved
current $j^{\mu}$ satisfying the continuity equation

\begin{equation}
  \label{eq:continuityequation}
  \partial_{\mu}j^{\mu}
  =
  0\,.
\end{equation}

\noindent
The amount of conserved charge enclosed in a spatial volume $\Sigma^{3}$
is defined as

\begin{equation}
  q
  \equiv
  \int_{\Sigma^{d-1}}d^{3}\Sigma_{\mu}j^{\mu}
  =
  \int_{\Sigma^{d-1}}d^{d-1}\Sigma n_{\mu}j^{\mu}\,,
\end{equation}

\noindent
where $n^{\mu}$ is the timelike unit vector normal to $\Sigma^{d-1}$. Then,
the continuity equation implies that the variation in time of this charge
equals the flux of this current across the boundary of the volume. In other
words: the amount of charge in that volume at a given time equals the initial
amount of charge plus or minus the amount of charge that entered or left the
volume.  The total charge of the Universe, then, must remain constant, because
nothing can enter or leave it from outside (\textit{i.e.}~there are no sources
nor sinks at infinity). We may characterize the Universe by the value of this
charge.

It is this property (which is what we mean by \textit{charge conservation})
that makes conserved charges real entities.

The continuity equation can be rewritten in a very convenient way in
differential-form language, using the $(d-1)$-form
$\mathbf{J}\equiv \star \left(j_{\mu}dx^{\mu}\right)$, which we will keep on
calling \textit{current}:

\begin{equation}
  \label{eq:continuityequation2}
  d\mathbf{J}
  =
  0\,.
\end{equation}

The charge contained in the spatial volume $\Sigma^{d-1}$ is now

\begin{equation}
  q
  =
  \int_{\Sigma^{d-1}}\mathbf{J}\,,
\end{equation}

\noindent
and charge conservation follows from Eq.~(\ref{eq:continuityequation2}) and
Stokes theorem: if $\Sigma^{d}$ is the region of spacetime swept by the
spatial volume $\Sigma^{d-1}$ in a given time and its boundary is

\begin{equation}
  \partial\Sigma^{d}
  =
  \Sigma^{d-1}_{t_{0}}\cup\Sigma^{d-1}_{t_{1}}\cup \mathcal{B}_{t_{0},t_{1}}\,,
\end{equation}

\noindent
integrating $d\mathbf{J}$ we find, taking into account the orientation of the
different pieces

\begin{equation}
  0
  =
  \int_{\Sigma^{d}}d\mathbf{J}
  =
\int_{\Sigma^{d-1}_{t_{1}}}\mathbf{J}  
-\int_{\Sigma^{d-1}_{t_{0}}}\mathbf{J}
+\int_{\mathcal{B}_{t_{0},t_{1}}}\mathbf{J}\,,  
\end{equation}

\noindent
which leads to

\begin{equation}
  q_{t_{1}}
  =
  q_{t_{0}} -\text{flux}_{t_{0}}^{t_{1}}\,,
\end{equation}

\noindent
where the flux

\begin{equation}
  \text{flux}_{t_{0}}^{t_{1}}
  \equiv
  \int_{\mathcal{B}_{t_{0},t_{1}}}\mathbf{J}\,,
\end{equation}

\noindent
is positive when the charge is leaving the volume. 

Let us now consider the charges contained in the volume outside a black
hole\footnote{Extending the integral to the interior has many practical and
  conceptual problems.}  at a given time $t_{0}$ and at a later time
$t_{1}$.\footnote{Here we must use Kruskal-Szekeres time, or other time
  coordinate which is well defined at the horizon, unlike Schwarzschild's.}
The difference between these charges is (minus) the amount of charge that has
crossed the horizon into the black hole. However, as a rule, black holes
cannot carry charges whose conservation follows from global symmetries and,
the net effect would be that the total amount of charge of the Universe would
decrease. This kind of charges which are always computed via volume integrals
cannot be used to characterize neither the black hole nor the whole black-hole
spacetime. Furthemore, there are indications that Quantum Gravity may break
all exact global symmetries. \cite{Banks:1988yz}.

In some cases, the $(d-1)$-form currents are related on-shell or
off-shell to $(d-2)$-form \textit{charges}\footnote{It is
  customary to use the same name for $q$ and for $\mathcal{Q}$. This does not
  lead to confusion, usually.} by

\begin{equation}
  \mathbf{J}
  =
  d\mathbf{Q}\,.
\end{equation}

Sometimes, this relation only holds asymptotically. This is, actually, what
happens with the gravitational field itself \cite{Abbott:1981ff}.  This
relation presents us with new opportunities: one can replace volume integrals
of $\mathbf{J}$ by surface integrals of $\mathbf{Q}$ over the boundaries
or, in general, over closed $(d-2)$-suerfaces $\Sigma^{d-2}$

\begin{equation}
  q
  =
  \int_{\Sigma^{d-2}}\mathbf{Q}\,.
\end{equation}

In black-hole spacetimes one can, then, do away with the problem of
integrating inside the horizon and decree that the total charge of the
spacetime is the integral of $\mathcal{Q}$ over the $(d-2)$-sphere at spatial
infinity $\Sigma^{d-2}=S^{d-2}_{\infty}$. Gravitational charges are defined in
this way and, as it is well known, it makes no sense to compute them
integrating anywhere else because, due to the principle of Equivalence, it is
not possible to define an energy/mass or momentum density function. Since
there is no time evolution at spatial infinity, these charges characterize the
whole spacetime.

In some cases the conserved charges obey ``Gauss laws'' because, at least in
the spacetime region we are concerned with, $\mathbf{J}=0$ and the
$(d-2)$-form $\mathbf{Q}$ is closed, typically on-shell\footnote{We use
  $\doteq$ to denote identities which are only satisfied on-shell.}

\begin{equation}
  d\mathbf{Q}
  \doteq
  0\,.
\end{equation}

\noindent
$(d-2)$-forms (and forms of other ranks) that satisfy this equation are often
called conserved charges as well, although this equation is not equivalent to
the continuity equation~(\ref{eq:continuityequation2}). Strictly speaking,
this equation is just the differential expression of the Gauss law: due to the
on-shell closedness of the $(d-2)$-form charge $\mathbf{Q}$, the value of the
integral does not change under smooth deformations of the integration
surfaces: if $\Sigma^{d-2\, \prime}$ has been obtained from $\Sigma^{d-2}$ by
a smooth deformation (which includes the assumption that the deformation does
not cross any point at which the classical equations of motion are not
satisfied) and $\Sigma^{d-1\, \prime}$ is the cobordant volume

\begin{equation}
  \partial\Sigma^{d-1\, \prime}
  =
  \Sigma^{d-2}\cup \Sigma^{d-2\, \prime}\,,
\end{equation}

\noindent
taking into account the orientations of $ \Sigma^{d-2}$ and
$\Sigma^{d-2\, \prime}$, Stokes theorem implies

\begin{equation}
  \int_{\Sigma^{d-2}}\mathbf{Q}-\int_{\Sigma^{d-2\, \prime}}\mathbf{Q}
  =
  \int_{\Sigma^{d-1\, \prime}}d\mathbf{Q}
  \doteq
  0\,,
\end{equation}

\noindent
which is the generalized form of the standard Gauss law of electromagnetism.

Gauss laws are typically satisfied by charges that source Abelian (uncharged)
fields, such as the electric charge of standard electromagnetism. In those
cases there is a very clear distinction between the sources of the fields and
the fields themselves. In general, the charges of gravitational or non-Abelian
Yang--Mills fields do not satisfy Gauss laws and only their values in the
whole spacetime can be defined as we explained before. For instance, since the
gravitational field carries energy, it is also the source of gravitational
field and any smooth deformation of the integration surface will change the
sources of gravitational field and consequently the value of the energy/mass
enclosed by it. This implies that there can be no closed $(d-2)$-form charge
describing a local energy density, as we mentioned before.

In spite of the above general discussion, there are special situations in
which it is actually possible to define on-shell-closed $(d-2)$-form charges
in theories of gravity \cite{Barnich:2003xg}. The main example is that of
spacetimes admitting a Killing vector $k$ in General Relativity without
matter.\footnote{More general on-shell and off-shell possibilities have been
  explored in Ref.~\cite{Gomez-Fayren:2023qly}.} The associated
on-shell-closed $(d-2)$-form is the \textit{Komar charge} $\mathbf{K}[k]$
\cite{Komar:1958wp}, which, in our notation and conventions,\footnote{Our
  conventions are those of Ref.~\cite{Ortin:2015hya}. In particular, we use
  mostly minus signature and we will always describe the gravitational field
  through the Vielbein $e^{a}=e^{a}{}_{\mu}dx^{\mu}$, where Latin indices are
  tangent-space indices and Greek indices are coordinate-basis
  indices. Furthermore, our Levi--Civita spin connection
  $\omega^{a}{}_{b}= \omega_{\mu}{}^{a}{}_{b}dx^{\mu}$ and its curvature
  2-form
  $R^{a}{}_{b}=\tfrac{1}{2}R_{\mu\nu}{}^{a}{}_{b}dx^{\mu}\wedge dx^{\nu}$ are
  defined through the relations
  \begin{subequations}
    \begin{align}
      \mathcal{D}e^{a}
      & \equiv
        de^{a}-\omega^{a}{}_{b}\wedge e^{b}
        =
        0\,,
      \\
      & \nonumber \\
      R^{a}{}_{b}
      & \equiv
        d\omega^{a}{}_{b}
        -\omega^{a}{}_{c}\wedge \omega^{c}{}_{b}\,,
    \end{align}
  \end{subequations}
  where $\mathcal{D}$ is the exterior Lorentz-covariant derivative.
  
  Equivalent ways of writing this $(d-2)$-form (ignoring the overall
  $(16\pi G_{N}^{(d)})$ factor) using the levi-Civita affine connection and
  its associated covariant derivative $\nabla$ are
  \begin{equation}
    \label{eq:Pintegrand}
    \begin{aligned}
      (-1)^{d-1}\star (e^{a}\wedge e^{b})P_{k\, ab}
      & =
        (-1)^{d-1}\frac{\varepsilon_{\mu_{1}\cdots \mu_{d-2}\alpha\beta}}{(d-2)!\,
        \sqrt{|g|}} \nabla^{\alpha}k^{\beta} 
        dx^{\mu_{1}}\wedge \cdots \wedge dx^{\mu_{d-2}}
      \\
      & \\
      & =
        d^{d-2}\Sigma_{\alpha\beta} \nabla^{\alpha}k^{\beta}
      \\
      & \\
      & =
        -d^{d-2}\Sigma\, n_{\alpha\beta} \nabla^{\alpha}k^{\beta}\,,
    \end{aligned}
  \end{equation}
  where $n_{\alpha\beta}$ is the binormal to the horizon, normalized as
  $n_{\mu\nu}n^{\mu\nu}=-2$.

  More details on our notation and conventions can be found in
  Ref.~\cite{Ortin:2015hya}.  } takes the form

\begin{equation}
  \label{eq:Komarcharge}
  \mathbf{K}[k]
  =
  (-1)^{d-1}\frac{1}{16\pi G_{N}{}^{(d)}}\star (e^{a}\wedge e^{b})P_{k\, ab}\,,
\end{equation}

\noindent
where $P_{k\, ab}$ is the \textit{Lorentz momentum map} or \textit{Killing
  bivector}

\begin{equation}
  P_{k\, ab}
  =
  \nabla_{a}k_{b}
  =
  \nabla_{[a}k_{b]}\,.
\end{equation}

In vacuum,

\begin{equation}
  d\mathbf{K}[k]
  \doteq
  0\,.
\end{equation}

The integral of this $(d-2)$-form charge over a $(d-2)$-sphere at spatial
infinity $S^{d-2}_{\infty}$ (often called \textit{Komar integral}) gives the
conserved gravitational charge associated with the Killing vector $k$, up to
normalization. For instance, in stationary spacetimes, the Komar integral
associated to the timelike Killing vector $\partial_{t}$ gives the mass of the
spacetime

\begin{equation}
  \label{eq:Komarmass}
\int_{S^{d-2}_{\infty}} \mathbf{K}[\partial_{t}]
  =
  \frac{d-3}{d-2}M\,,
\end{equation}

\noindent
while the Komar integral associated to the Killing vector that generates
rotations around some symmetry axis, $\partial_{\varphi}$, gives the component
of the angular momentum in that direction

\begin{equation}
  \label{eq:Komarangularmomentum}
\int_{S^{d-2}_{\infty}}  \mathbf{K}[\partial_{\varphi}]
  =
  J\,.
\end{equation}

Since the Komar charge is closed, the same results can be obtained by
integrating over any other $(d-2)$ surface that can be smoothly taken to
infinity. If one chooses spheres, the integrands are independent of the radial
coordinate. In particular, in stationary black-hole spacetimes one can
integrate over any section of the event horizon obtaining exactly the same
results. Different choices of section correspond to different choices of
hypersurface $\Sigma^{d-1}$.

What makes this seemingly trivial result really interesting is that the
integral over the horizon is naturally expressed in terms of the physical
variables associated to the horizon, such as the Hawking temperature $T$ and
Bekenstein-Hawking entropy $S$, giving rise to an identity (the \textit{Smarr
  formula} \cite{Smarr:1972kt}) that relates, in a highly non-trivial way
these two sets of physical variables \cite{Bardeen:1973gs,Carter:1973rla,
  Magnon:1985sc,Bazanski:1990qd,Kastor:2008xb,Kastor:2010gq,Banerjee:2010yd,Banerjee:2010ye,Ortin:2021ade}.\footnote{Other
methods have been used to derive Smarr formulas. For instance, in
Ref.~\cite{Heusler:1997am} the reduction of the action to that of a
$\sigma$-model was exploited to this end. See also Ref.~\cite{Chrusciel:2012jk}.}

Let us show how the Smarr formula can be obtained in this way.

If the horizon $\mathcal{H}$ is non-degenerate, it is most convenient to
choose the bifurcation surface $\mathcal{BH}$ because it leads to great
simplifications. In particular, if
$k=\partial_{t}-\Omega_{H}\partial_{\varphi}$ is the Killing vector that
becomes null on it, $k^{2}\stackrel{\mathcal{H}}{=}0$, one can show that, on
the bifurcation surface $\mathcal{BH}$

\begin{equation}
  P_{k\, ab}
  \stackrel{\mathcal{BH}}{=}
  \kappa n_{ab}\,,
\end{equation}

\noindent
where $\kappa$ is the surface gravity.  Then, using Eq.~(\ref{eq:Pintegrand})
and the zeroth law of black-hole mechanics
($d\kappa \stackrel{\mathcal{H}}{=}0$) \cite{Bardeen:1973gs} we get,
integrating over the bifurcation surface

\begin{equation}
  \int_{\mathcal{BH}}  \mathbf{K}[k]
  =
  -\frac{\kappa}{16\pi G_{N}^{(d)}} \int_{\mathcal{BH}}d^{(d-2)}\Sigma\, 
  n_{ab}n^{ab}
  =
  \frac{\kappa A_{H}}{8\pi G_{N}^{(d)}}
  =
  TS\,.
\end{equation}

On the other hand, using Eqs.~(\ref{eq:Komarmass}) and
(\ref{eq:Komarangularmomentum}) the integral at spatial infinity gives

\begin{equation}
\int_{S^{d-2}_{\infty}}  d\mathbf{K}[k]
  =
  \int_{S^{d-2}_{\infty}} \mathbf{K}[\partial_{t}]
  -\Omega_{H}\int_{S^{d-2}_{\infty}}\mathbf{K}[\partial_{\varphi}]
  =
  \frac{d-3}{d-2}M -\Omega_{H}J\,,
\end{equation}

\noindent
where we have also used the fact that $\Omega_{H}$ is constant over the
horizon, which can be understood as a generalization of the zeroth law.

The on-shell-closedness of the Komar charge tells us that the results of these
two integrals are equal

\begin{equation}
  \label{eq:pureGRSmarrformula}
M = \frac{d-2}{d-3}\left( TS +\Omega_{H}J\right)\,,
\end{equation}

\noindent
which is the Smarr formula for stationary, asymptotically flat black holes in
vacuum.

The Komar charge Eq.~(\ref{eq:Komarcharge}) is no longer closed on-shell when
gravity is coupled to most forms of matter and the above algorithm cannot be
used directly to derive Smarr formulas. One can follow the procedure outlined
in Ref.~\cite{Townsend:1997ku} (see, for instance,
Ref.~\cite{Herdeiro:2024pmv}) involving volume integrals but these do not give
directly expressions involving the total, conserved ADM charges. The
derivation of the Smarr formulas is, therefore, not as clean as the one
puoneered by Bardeen, Carter and Hawking \cite{Bardeen:1973gs,Carter:1973rla}
that we have just explained here.

The Komar charge of General Relativity in absence of matter is just
(minus)\footnote{In our conventions.} the Noether $(d-2)$-form charge
associated to the invariance of the theory under diffeomorphisms (also known
as \textit{Noether--Wald charge}) $\mathbf{Q}[\xi]$ evaluated on a Killing
vector $k$ which leaves invariant the only field of this theory, the metric,
that is

\begin{equation}
  \label{eq:invarianceofthemetric}
  \delta_{k}g_{\mu\nu}
  =
  -\pounds_{k}g_{\mu\nu}
  =
  0\,.
\end{equation}

\noindent
where $\pounds_{k}$ is the Lie derivative.

This suggests that (minus) the Noether--Wald charge of the matter-coupled
General Relativity theory, $-\mathbf{Q}[\xi]$, evaluated over vector fields
$k$ generating symmetries of all the fields of the theory that we denote
generically by $\varphi$,

\begin{equation}
  \delta_{k}\varphi
  =
  0\,,
\end{equation}

\noindent
may provide the on-shell-closed generalization of the Komar charge
Eq.~(\ref{eq:Komarcharge}) that we need. Observe that the above condition
implies Eq.~(\ref{eq:invarianceofthemetric}) and $k$ of a Killing vector of
the spacetime metric.

However, as observed in Ref.~\cite{Liberati:2015xcp} (see also
Ref.~\cite{Golshani:2024fry}), in most diffeomorphism-invariant
theories\footnote{In theories with Chern--Simons terms, invariant under gauge
  transformations up to total derivatives, there could be additional terms in
  the right-hand side of this expression. This is due to the fact, to be
  explained shortly, that diffeomorphisms induce gauge transformations.}

\begin{equation}
  d\mathbf{Q}[k]
  \doteq
  \imath_{k}\mathbf{L}\,,
\end{equation}

\noindent
where $\mathbf{L}$ is the $d$-form Lagrangian and $\imath_{k}\mathbf{L}$ is
its interior product with the Killing vector field $k$. Integrating both sides
of this equation over a hypersurface $\Sigma^{d-1}$ with boundaries at
infinity and at the bifurcation surface

\begin{equation}
  \label{eq:favoritehypersurfaces}
  \partial \Sigma^{d-1}
  =
  S^{d-2}_{\infty}  \cup \mathcal{BH}\,,
\end{equation}

\noindent
and applying Stokes theorem

\begin{equation}
  \int_{\Sigma^{d-1}} d\mathbf{Q}[k]
  =
\int_{S^{d-2}_{\infty}} \mathbf{Q}[k]
-\int_{\mathcal{BH}} \mathbf{Q}[k]
  \doteq
  \int_{\Sigma^{d-1}}\imath_{k}\mathbf{L}\,.
\end{equation}

This relation can be used to obtain generalized Smarr formulas
\cite{Liberati:2015xcp},\footnote{See also the more recent
  Ref.~\cite{Golshani:2024fry}, based on the results of
  Ref.~\cite{Adami:2024gdx}.} but the volume integral term obscures their
interpretation as relations between physical quantities defined at the horizon
and spatial infinity.

In Ref.~\cite{Ortin:2021ade} it was argued that, as long as the diffeomorphism
generated by $k$ leaves invariant all the fields of the theory\footnote{When
  the theory contains fields with gauge freedoms, the transformations
  generated by $k$, $\delta_{k},$ acting on them have to be defined carefully
  so that the statements $\delta_{k} \varphi=0$ are gauge invariant
  \cite{Elgood:2020svt,Elgood:2020mdx,Elgood:2020nls}. This point will be
  explained in more detail shortly.}  for a given solution, and, hence, leaves
invariant the Lagrangian evaluated on-shell, $\imath_{k}\mathbf{L}$ is an
exact $(d-1)$-form. Indeed,

\begin{equation}
  0
  \doteq
  \delta_{k}\mathbf{L}
  =
  -\pounds_{k}\mathbf{L}
  =
  -d\imath_{k}\mathbf{L}\,,
\end{equation}

\noindent
implies the local existence of a $(d-2)$-form $\omega_{k}$ such that  
\begin{equation}
  \label{eq:ikL=domegak}
  \imath_{k}\mathbf{L}
  \doteq
  d\omega_{k}\,,
\end{equation}

\noindent
and we can define the generalized Komar $(d-2)$-form charge

\begin{equation}
  \label{eq:generalizedKomarchargedef}
  \mathbf{K}[k]
  \equiv
  -\left(\mathbf{Q}[k]-\omega_{k}\right)\,,
\end{equation}

\noindent
which is closed on-shell by construction.

In many
\cite{Mitsios:2021zrn,Meessen:2022hcg,Ortin:2022uxa,Ballesteros:2023iqb,Bandos:2023zbs}
but not all \cite{Ballesteros:2023muf} theories it is possible to give an
explicit expression of $\omega_{k}$ which just needs to be evaluated on a
given solution. In all the cases we have studied so far the algorithm gives a
Smarr formula with terms which are products of a thermodynamic charge $q$
times its conjugate chemical potential $\Phi$. These objects appear in the
first law in terms of the form $\Phi\delta q$ and sometimes they have to be
understood in the context of extended thermodynamics.

The use of the generalized Komar charge is not necessarily restricted to the
derivation of Smarr formulas for black holes.\footnote{Scalar charges
  satisfying Gauss laws have also been used to prove no-hair theorems in
  Ref.~\cite{Ballesteros:2023iqb} (see Section~\ref{sec-Estheory}) and to
  prove that certain initial data sets do not correspond to constant-time
  slices of stationary black-hole solutions in
  Ref.~\cite{Vinckers:2024tsa}. We will use these and other charges in our
  analysis.} Here we want to use it to study stationary, globally regular,
horizonless, topologically trivial, asymptotically flat solutions of General
Relativity coupled to bosonic fields that we will loosely call ``boson
stars'', although this name is used in a slightly more restricted sense in the
literature (see, for instance, Refs.~\cite{Liebling:2012fv,Herdeiro:2024pmv}
and references therein). These solutions can be physically understood as
self-gravitating solitons of some bosonic fields whose density is not high
enough to cause gravitational collapse.

As a simple example of what we intend to do, let us consider possible boson
star solutions of General Relativity in vacuum. The boundary of any Cauchy
hypersurface of these spaces is the $(d-2)$-sphere at infinity. Integrating
$d\mathbf{K}[\partial_{t}]\doteq 0$ over the Cauchy hypersurface and using
Stokes theorem and Eq.~(\ref{eq:Komarmass}) leads to $M=0$ so that, according
to the positive mass theorem \cite{Schon:1979rg,Schon:1981vd,Witten:1981mf},
the solution must be Minkowski spacetime. In other words: there are no boson
stars in vacuum.

This example shows how the generalized Komar charge can be used to constrain
the possible solutions of a given theory.\footnote{The study of the
  (non-)extsitence of non-topological solitons has a long history. The first
  result of this kind is Derrick's theorem Ref.~\cite{Derrick:1964ww} which
  proves the non-extstence of particle-like solutions of non-linear
  generalizations of the wave equation. The possibility of the existence of
  those solutions if they were allowed to have a periodic time dependence was
  also suggested in that reference. Solutions of the Einstein-Klein-Gordon
  theory of this kind were first found by Kaup in Ref.~\cite{Kaup:1968zz}. The
  characterization of non-gravitating spherically-symmetric solitons (called
  there \textit{Q-balls}) by the charge associated to a non-gauged global
  symmetry (a U$(1)$ symmetry, typically) and their existence were studied by
  Coleman in Ref.~\cite{Coleman:1985ki}. It is this symmetry that is used to
  make a generalized symmetric ansatz, as we will explain. Stars and black
  holes with solitonic scalar fields having a U$(1)$ global symmetry were
  constructed by Friedberg, Lee and Pang in Ref.~\cite{Friedberg:1986tq}.} It
also tells us that, in order to make non-trivial solutions possible we have to
relax one or several of these assumptions: stationarity, regularity,
asymptotic flatness, topological triviality, absence of event horizons and
Einstein equations in vacuum.

We have already seen that a non-degenerate event horizon works as a second,
inner boundary\footnote{This is not completely true: we can always choose a
  hypersurface $\Sigma^{d-1}$ that goes inside the horizon. However, in
  general, inside the horizon we are going to find singularities or
  non-trivial topology, including other asymptotically-flat regions. If we do
  not want to deal with these, we must consider another boundary and a section
  of the horizon (its bifurcation surface, if any) is the most convenient
  place for it due to its special properties.} on which the Komar integral of
the Killing vector that generates it gives $TS$, so that
$M-\Omega_{H}J =2TS \geq 0$, which can be satisfied for $M\neq 0$, even if
the horizon is degenerate and $T=0$.

In absence of horizons, for obvious reasons, we do not want to give up on the
regularity of the solutions and, since we are interested in stationary,
asymptotically-flat solutions of trivial topology, we can only play with two
assumptions: the coupling to different kinds of matter (so we do not deal with
the vacuum Einstein equations) and with the implementation of the stationarity
assumption on the matter fields, which can modify the Komar charge. Most
often, we deal with axisymmetric spacetimes. The condition of axisymmetry can
also be implemented in different ways.

This requires an explanation.

An asymptotically-flat spacetime is called stationary when it admits a Killing
vector field which is asymptotically timelike and axisymmetric when it admits
a Killing field vector with periodic orbits and fixed points (the axis). Let
us call $k$ any of these vector fields. The conditions of invariance are
expressed by Eq.~(\ref{eq:invarianceofthemetric}) above.  If this metric is
part of a solution that includes matter fields it is commonly assumed that
they should also be exactly invariant under the diffeomorphism generated by
$k$ as well. This is usually expressed as

\begin{equation}
  \label{eq:symmetricansatz}
  \delta_{k}\varphi
  =
  -\pounds_{k}\varphi
  =
  0\,.
\end{equation}

Most theories are invariant under local and global symmetries which act on the
same fields and we must take this fact into account in the above
expression. Let us start with gauge symmetries.

\subsection{Spacetime symmetries of fields with gauge freedoms}

In general, gauge symmetries do not commute with the standard Lie derivative
and the above expression is, at the very least, ambiguous, because it is not
gauge-invariant. Using it can lead to wrong results \cite{Elgood:2020nls} and,
as repeatedly argued in
Refs.~\cite{Elgood:2020svt,Elgood:2020mdx,Elgood:2020nls} it must be replaced
by a \textit{gauge-covariant Lie derivative} $\mathbb{L}_{k}$ that annihilates
the fields in a gauge-invariant way for the Killing vector $k$. This covariant
Lie derivative is always a combination of the standard one
$\pounds_{k}\varphi$ and a field- and Killing-vector-dependent gauge
transformation\footnote{The field-dependent parameters of these gauge
  transformations must be the same for all the fields transforming under the
  same symmetry, though.} of the field
$\delta_{\Lambda(k,\varphi)}\varphi$. Thus, we are led to use the
transformation

\begin{equation}
  \delta_{k}\varphi
  =
  -\mathbb{L}_{k}\varphi
  =
  -\left(\pounds_{k}-\delta_{\Lambda(k,\varphi)}\right)\varphi
  =
  0\,.
\end{equation}

The gauge transformation $\delta_{\Lambda(k,\varphi)}$ can be seen as
``induced'' by the diffeomorphism generated by the Killing vector.

An alternative way of looking at this transformation is to consider that the
best we can do with field with gauge freedoms is to ask for invariance under
diffeomorphisms up to gauge transformations

\begin{equation}
  \label{eq:diffandcompensatinggauge}
  \pounds_{k}\varphi
  =
  \delta_{\Lambda(k,\varphi)}\varphi\,.  
\end{equation}

While the first point of view is actually closer to the rigorous description
of the action of diffeomorphisms in fiber bundles,\footnote{See,
  \textit{e.g.}~\cite{Prabhu:2015vua,Bunk:2023ojb}.} the second is going to
help us to understand better the implementation of the stationarity condition
on fields transforming under some global symmetries of the theory.

\subsection{Spacetime symmetries of fields with global freedoms}

Let us denote the action of the independent global symmetries by
$\delta_{I}\varphi$ where $I,J,K,\ldots$ label them. Indeed, looking at the
above transformation, it is very natural to generalize the standard
\textit{symmetric} ansatz Eq.~(\ref{eq:symmetricansatz}) to\footnote{For
  simplicity we consider only gauge-invariant fields.}

\begin{equation}
  \label{eq:generalizedsymmetryansatz0}
  \pounds_{k}\varphi
  =
  \vartheta_{k}{}^{I}\delta_{I}\varphi\,,
\end{equation}

\noindent
where the $\vartheta_{k}{}^{I}$ are constants that depend on the Killing
vector $k$\footnote{Here we are considering just one isometry and the index
  $k$ in $\vartheta_{k}{}^{I}$ is redundant but it will be relevant when we
  consider more general situations with several isometries in
  Appendix~\ref{sec-generalconsiderations}.} and which must be compatible
with the global structure (the periodicity, for instance, see
footnote~\ref{foot:esafootnote} in page~\pageref{foot:esafootnote}) of the
symmetries involved.

For scalar fields, for instance, the above \textit{generalized symmetric}
ansatz allows dependence on the coordinate adapted to the isometry generated
by $k$, in contrast to what happens to the metric. It can be shown, however,
that the generalized symmetric ansatz leads to exactly symmetric
energy-momentum tensors $\pounds_{k}T_{\mu\nu}=0$ \cite{Yazadjiev:2024rql} so
that it is perfectly compatible with the isometry. More generally, if $T$ is
any tensor invariant under the global symmetry,

\begin{equation}
  \pounds_{k}T
  =
  \vartheta_{k}{}^{I}\delta_{I}T
  =
  0\,.
\end{equation}

There is an important point concerning the generalized symmetry ansatz
Eq.~(\ref{eq:generalizedsymmetryansatz0}) that we would like to clarify here.
Most of the General Relativity literature considers that the generalized
symmetric ansatz for the scalar fields prevents the spacetime isometries from
becoming true symmetries of the complete field configuration. A common way of
expressing this idea is by saying that the scalar fields ``do not inherit''
the symmetry of the spacetime metric.\footnote{See, for instance
  Ref.~\cite{Smolic:2015txa} and references therein.} However, it follows from
our previous discussion of the action of isometries on fields with gauge
symmetries that they do not ``inherit'' spacetime symmetries in the usual,
restricted, sense in which spacetime symmetries act on all fields through
standard Lie derivatives. Still, there is a perfectly well defined (and
gauge-invariant) sense in which the diffeomorphism generated by the vector $k$
leaves the field invariant, expressed through the combination of the standard
Lie derivative and the ``compensating'' gauge transformation into the
gauge-covariant Lie derivative. Gauge fields may be time-dependent in a
certain gauge and, still, be invariant under the gauge-covariant Lie
derivative with respect to $\partial_{t}$ so that this isometry is a symmetry
of all the fields of the theory and not just of the metric. The solution is
stationary even if the gauge fields have dependence on $t$. Similar remarks
apply to axisymmetry and the dependence on the angular coordinate $\varphi$

The same reasoning can be used in the case of global symmetries: the fields
are now invariant under the transformation

\begin{equation}
  \delta_{k}\varphi
  \equiv
  -\left(\pounds_{k}
  -\vartheta_{k}{}^{I}\delta_{I}\right)\varphi
  =
  0\,,
\end{equation}

\noindent
so that $\delta_{k}$ is a symmetry of all the fields of the theory, although
it does not simply act as the Lie derivative with respect to $k$ on all the
fields.  For $k=\partial_{t}$, each possible choice of the constants
$\vartheta_{k}{}^{I}$ provides a different implementation of the stationarity
condition but for any choice, there is a well-defined sense in which we can
still say that the solution is stationary, even if the matter fields have a
dependence on $t$.

All the boson star solutions found so far are based on non-trivial choices of
$\vartheta_{k}{}^{I}$s for $\partial_{t}$ and $\partial_{\varphi}$ in
stationary and axisymmetric spacetimes.  One of our goals is to understand
this fact using the generalized Komar charge which must be modified by those
choices.

In this paper we are going to consider different kinds of matter coupled to
gravity and different implementations of the stationarity and axisymmetry
ansatzs on them, constructing the generalized Komar charge and other charges
satisfying Gauss laws in each case and finding the implications for the
existence of horizonless, globally regular, topologically trivial,
asymptotically flat solutions in $d=4$ dimensions.  It is organized as
follows: In Section~\ref{sec-Estheory} we consider the simplest kind of
matter, namely a single, real, massless scalar field. Next, in
Section~\ref{sec-EMtheory} we consider the Einstein--Maxwell theory. In
Section~\ref{sec-Estheoryscalarpotential} we add a scalar potential to the
theory of Section~\ref{sec-Estheory}, breaking its shift symmetry.  In
Section~\ref{sec-EMstheories} we consider theories with an arbitrary number of
scalars and Abelian vector fields which have the generic form of a
4-dimensional, ungauged supergravity theory, combining and generalizing the
results of the previous sections. Section~\ref{sec-discussion} contains a
discussion of our results. Appendix~\ref{sec-generalconsiderations} contains a
general discussion of the generalized symmetric ansatz using only scalar
fields as an example: consistency conditions (very similar to the famous
quadratic constraint of the embedding tensor formalism) and the definition of
scalar charges. Appendix~\ref{sec-generalizedzerothlaw} reviews the proof of
the generalized zeroth law for electrostatic and magnetostatic potentials that
we use in the main text. Finally, in Appendix~\ref{sec-trial} we show that
solutions satisfying the generalized symmetric ansatz with respect to
electric-magnetic duality rotations (a symmetry of the equations of motion but
not of the action) can be found, even though they turn out not to be very
interesting in this simple case.

\section{The Einstein--scalar theory}
\label{sec-Estheory}

The simplest kind of matter that can be coupled to Einstein's gravitational
field which we will always describe through the Vierbein
$e^{a}=e^{a}{}_{\mu}dx^{\mu}$, is a real, massless scalar $\phi$. The action
for this Einstein--scalar (ES) theory is

\begin{equation}
\label{eq:Esaction}
  S[e,\phi]
   =
    \frac{1}{16\pi G_{N}^{(4)}}\int 
    \left[ -\star (e^{a}\wedge e^{b}) \wedge R_{ab}
    +\tfrac{1}{2}d\phi\wedge \star d\phi
    \right]
 \equiv
    \int \mathbf{L}\,.
\end{equation}

If the scalar field satisfies the standard symmetric ansatz
Eq.~(\ref{eq:symmetricansatz}) for the timelike Killing vector
$k=\partial_{t}$ it is not difficult to see that the generalized Komar charge
of this theory is equal to the standard one given in
Eq.~(\ref{eq:Komarcharge}) (see, for instance,
Ref.~\cite{Ballesteros:2023muf}). This implies that there are no boson star
solutions\footnote{As discussed in the Introduction, we are using here the
  term ``boson star'' in a rather loose way. This is probably more evident in
  this case and more conventional or proper way of expressing this result
  would be the non-existence of non-topological solitons \cite{Lee:1991ax}.}
in this theory with the standard implementation of the stationarity
condition. On the other hand, according to the Smarr formula
Eq.~(\ref{eq:pureGRSmarrformula}), the presence of a non-degenerate horizon
allows for black holes of mass $M=2ST+2\Omega_{H}J$.

Before we study whether the non-trivial implementations of the stationarity
and axisymmetry conditions through the generalized symmetric ansatz can modify
this conclusion, it is convenient to define a charge for the scalar $\phi$
that satisfies a Gauss law in a stationary spacetime
\cite{Pacilio:2018gom,Ballesteros:2023iqb}. Since the theory is invariant
under global shifts of the scalar field, there is an on-shell-conserved
Noether current whose Hodge-dual 3-form we denote by $J_{\phi}$ and which is
given by

\begin{equation}
  \label{eq:EstheoryJ}
  J_{\phi}
  \equiv
  \frac{1}{16\pi G_{N}^{(4)}} \star d\phi\,.
\end{equation}

\noindent
The on-shell conservation is here equivalent to the on-shell-closedness of
$J_{\phi}$:

\begin{equation}
  \label{eq:dJdoteq0}
  dJ_{\phi}
  =
  -\mathbf{E}_{\phi}
  \doteq
  0\,,
\end{equation}

\noindent
where $\mathbf{E}_{\phi}$ is the equation of motion of $\phi$.

The charge associated to the above 3-form, given by its integral over a
3-dimensional spacelike hypersurface,\footnote{This is equivalent to the
  volume integral of the timelike component of the standard current.} though,
does not satisfy a Gauss law. Furthermore, in stationary spacetimes in which
the scalar is time-independent it vanishes identically
\cite{Ballesteros:2023iqb} because the timelike component of the current does.
A different charge is needed to characterize the scalar field.

If all the fields are exactly invariant under the diffeomorphism generated by
$k$, so is $J_{\phi}$ and, using Cartan's magic formula and Eq.~(\ref{eq:dJdoteq0})

\begin{equation}
  \label{eq:constructionofscalarcharges}
  0
  =
  \delta_{k}J_{\phi}
  =
  -\pounds_{k}J_{\phi}
  =
  -(\imath_{k}d+d\imath_{k})J_{\phi}
  \doteq
  d\imath_{k}J_{\phi}\,,
\end{equation}

\noindent
which tells us that the 2-form $\mathbf{Q}_{\phi}[k]=\imath_{k}J_{\phi}$ is
closed on-shell and the charge $\Sigma_{\phi}$ defined by its integral over
closed spacelike surfaces satisfies a Gauss law
\cite{Pacilio:2018gom,Ballesteros:2023iqb}. In standard spherical coordinates
$\Sigma$ appears as the coefficient of the $1/r$ term in the asymptotic
expansion of $\phi$.
 
The same arguments used for the Komar charge show that boson stars would be
characterized by $\Sigma_{\phi}=0$ which is consistent with their
non-existence.

In presence of a non-degenerate horizon $\mathcal{H}$, using the Killing
vector $k$ normal to it ($k^{2}\stackrel{\mathcal{H}}{=}0$) and integrating
$d\mathbf{Q}_{\phi}[k]=0$ over a hypersurface satisfying
Eq.~(\ref{eq:favoritehypersurfaces}), taking into account that, by definition,
$k\stackrel{\mathcal{BH}}{=}0$ so that
$\mathbf{Q}_{\phi}[k]\stackrel{\mathcal{BH}}{=}0$, we find again
$\Sigma_{\phi}=0$. This result can be interpreted as a no-hair theorem
relating the existence of a non-degenerate horizon to the absence of scalar
charge of the kind we have defined above \cite{Ballesteros:2023iqb}.

\subsection{The Einstein--scalar theory and the generalized symmetric ansatz}
\label{sec-EMtheorygeneralized}

Now we want to see how the previous results may be modified if we implement
the stationarity condition ($k=\partial_{t}$) in the
form\footnote{\label{foot:esafootnote}Observe that this ansatz is consistent
  because both the scalar field $\phi$ and the time coordinate $t$ take values
  in $\mathbb{R}$. In axisymmetric spacetimes, whose metric is invariant under
  constant shifts of the angular coordinate $\varphi\sim \varphi+2\pi$, the
  ansatz $\partial_{\varphi}\phi=\vartheta_{\varphi}$ leads to a multivalued
  scalar field unless it is assumed that $\phi$ must also be identified with
  $\phi+2\pi \vartheta_{\varphi}$. This is not possible in this theory, but in
  theories with more scalar fields that can be understood as coordinates in a
  ``target space'', some of them may also be understood as angular,
  periodically identified, coordinates and the ansatz would be consistent.
  This is the case of the phase of the complex Klein--Gordon scalar, for
  instance \cite{Herdeiro:2014goa,Herdeiro:2015gia}. We will consider this
  possibility in Section~\ref{sec-EMSgeneralizedsymmetricansatz}. }

\begin{equation}
  \label{eq:generalizedsymmetricansatzfreescalar}
  \delta_{k}\phi
  =
  -(\pounds_{k}\phi - \vartheta_{k})
  =
  0\,,
\end{equation}

\noindent
allowing for linear dependence of the scalar on the time coordinate:

\begin{equation}
  \phi
  =
  \vartheta_{k} t +f(x^{1},x^{2},x^{3})\,.
\end{equation}

Notice that at any given time $t=t_{0}$ it is possible to eliminate the
$\vartheta_{k} t_{0}$ using the shift symmetry. The term $\vartheta_{k} t$ is
only relevant globally.

In this setting, the Noether current associated to the invariance of the
theory under constant shifts of the scalar, $J_{\phi}$, defined in
Eq.~(\ref{eq:EstheoryJ}) does not vanish any longer.

The Noether--Wald charge of the theory, $\mathbf{Q}[\xi]$, is associated to
the invariance of the action under diffeomorphisms only and, therefore, it is
not modified by the generalized symmetric ansatz. It has the same expression
as in the vacuum theory, namely

\begin{equation}
  \mathbf{Q}[\xi]
  =
\frac{1}{16\pi G_{N}^{(4)}}  \star (e^{a}\wedge e^{b})P_{\xi\, ab}\,.
\end{equation}

By construction,

\begin{equation}
  d\mathbf{Q}[\xi]
  =
  \mathbf{\Theta}(e^{a},\phi,\delta_{\xi}e^{a}, \delta_{\xi} \phi)
  +\mathbf{E}_{a} \xi^{a}
  +\imath_{\xi}\mathbf{L}\,. 
\end{equation}

\noindent
where the presymplectic potential is given by

\begin{equation}
  \mathbf{\Theta}(e^{a},\phi,\delta_{\xi}e^{a}, \delta_{\xi} \phi)
  =
  \star (e^{a}\wedge e^{b})\wedge
  \left(\mathcal{D}P_{\xi}{}^{ab}+\imath_{\xi}R^{ab}\right)
   -\star d\phi \imath_{\xi}d\phi \,.
\end{equation}

When $\xi=k$ the term in parenthesis vanishes identically.\footnote{The
  equation
  \begin{equation}
    \mathcal{D}P_{k}{}^{ab}+\imath_{k}R^{ab}
    =
    0\,,
  \end{equation}
  defines the Lorentz momentum map $P_{k}{}^{ab}$ and is solved when
  $P_{k}{}^{ab}=\nabla_{a}k_{b}$, the Killing bivector. Actually, replacing
  the Lorentz momentum map by the Killing bivector this equation becomes the
  integrability condition of the Killing vector equation. On the other hand,
  the left-hand side of the equation is just $-\delta_{k}\omega^{ab}$
  \cite{Elgood:2020svt}.} With the standard symmetric ansatz, the second term
vanishes as well and, on-shell, we are left with
$d\mathbf{Q}[k] \doteq\imath_{k}\mathbf{L}$, but in this theory
$\mathbf{L} \doteq 0$ and the Komar charge coincides with (minus) the
Noether--Wald charge and with that of the vacuum theory.

In the generalized symmetric case
Eq.~(\ref{eq:generalizedsymmetricansatzfreescalar}) the Lagrangian and the
first term in the presymplectic potential still vanish on-shell, but the
second term does not and

\begin{equation}
  \mathbf{\Theta}(e^{a},\phi,\delta_{k}e^{a}, \delta_{k} \phi)
  =
  -\vartheta_{k} \star d\phi
  =
  -\vartheta_{k} J_{\phi}\,,
\end{equation}

\noindent
so that 

\begin{equation}
  d\mathbf{Q}[k]
  \doteq 
  -\vartheta_{k} J_{\phi}\,.
\end{equation}

Since $J_{\phi}$ is closed on-shell, locally there must exist a 2-form charge
$\mathbf{Q}_{\phi}$ such that

\begin{equation}
  J_{\phi}
  \doteq
  d\mathbf{Q}_{\phi}\,,
\end{equation}

\noindent
and we can define the on-shell-closed generalized Komar charge

\begin{equation}
  \mathbf{K}[k]
  \equiv
  -\left(\mathbf{Q}[k]+\vartheta_{k}\mathbf{Q}_{\phi}\right)\,,
\end{equation}

\noindent
which differs from the one obtained with the standard implementation of the
stationarity condition.

In this case there is no general expression for $\mathbf{Q}_{\phi}$: it
depends on the explicit form of the solution. Its actual form will not be
important in what follows, though.

Let us consider a boson star solution and let us integrate
$d\mathbf{K}[k] \doteq 0$ over a spacelike hypersurface bounded by a 2-sphere
of radius $r$. Applying Stokes theorem we find that

\begin{equation}
  \int_{S^{2}_{r}} \mathbf{Q}[k]
  \doteq 
  -\vartheta_{k} \int_{\Sigma^{2}_{r}} \mathbf{Q}_{\phi}\,.
\end{equation}

\noindent
The integral in the left-hand side converges to $M/2$ as $r$ approaches
infinity. The integral in the right-hand side gives the scalar charge
contained in the 2-sphere and it is easy to see that this charge diverges when
$r$ approaches infinity. This indicates that boson stars of this kind do not
exist in this theory. The situation is not improved by admitting the existence
of an event horizon or rotation since the integral that gives the charge keeps
diverging at infinity. Thus, stationary, asymptotically flat black-hole
solutions of this theory do not exist, either. Coupling the scalar field to
itself or to some other source can change these conclusions, as we are going
to see in Section~\ref{sec-Estheoryscalarpotential}.

For black-hole spacetimes there is another way to arrive at this result which
does not rely on these charges and which will play an important role later. If
$k$ is the Killing vector that becomes null on the horizon
$k^{2}\stackrel{\mathcal{H}}{=}0$ one can show that

\begin{equation}
   k^{\mu}k^{\nu}R_{\mu\nu}
  \stackrel{\mathcal{H}}{=}
  0\,,
\end{equation}

\noindent
which implies, upon use of the Einstein equations
and $k^{2}\stackrel{\mathcal{H}}{=}0$ 

\begin{equation}
  \label{eq:TkkHo}
   k^{\mu}k^{\nu}T_{\mu\nu}
  \stackrel{\mathcal{H}}{=}
  0\,,
\end{equation}

\noindent
leading, in this particular theory, to the conclusion

\begin{equation}
  \label{eq:ikdphihorizon}
  \imath_{k}d\phi
  \stackrel{\mathcal{H}}{=}  
  0\,,
\end{equation}

\noindent
so that $\vartheta_{k}=0$. If the black hole is rotating
$k=\partial_{t}-\Omega\partial_{\varphi}$ and, since in this case
$\partial_{\varphi}\phi=0$ (see footnote~\ref{foot:esafootnote} in
page~\pageref{foot:esafootnote}), we arrive at the same conclusion as
before. Notice, though, that in the cases in which it is consistent to impose
the generalized symmetric ansatz

\begin{subequations}
  \begin{align}
    \partial_{t}\phi
    & =
      \vartheta_{t}\equiv \omega\,,
    \\
    & \nonumber \\
    \partial_{\varphi}\phi
    & =
      \vartheta_{\varphi}\equiv m\,,
  \end{align}
\end{subequations}

\noindent
with non-vanishing $\omega$ and $m$, Eq.~(\ref{eq:ikdphihorizon}) demands

\begin{equation}
  \label{eq:synchronizationcondition}
  \frac{\omega}{m}
  =
  \Omega_{H}\,,
\end{equation}

\noindent
known as ``synchronization condition'' \cite{Herdeiro:2024pmv}, on account of
the relation $\vartheta_{k}=\omega -\Omega_{H}m$.

\section{The Einstein--Maxwell theory}
\label{sec-EMtheory}

The Maxwell (or electromagnetic) field provides $A=A_{\mu}dx^{\mu}$ another
simple kind of matter that can be coupled to Einstein's gravitational
field. It is worth stressing that, despite the notation, geometrically $A$ is
not a 1-form, but a gauge field, a connection in a U$(1)$ fiber bundle, with
gauge transformations

\begin{equation}
  \delta_{\chi}A
  =
  d\chi\,,
\end{equation}

\noindent
where $\chi$ is any real function. Its gauge-invariant 2-form field strength,
locally given by

\begin{equation}
  F
  \equiv
  dA\,,
\end{equation}

\noindent
is a 2-form, though.

The action of the 4-dimensional Einstein--Maxwell (EM) theory is

\begin{equation}
  \label{eq:EMaction}
    S[e,A]
    =
      \frac{1}{16\pi G_{\rm N}^{(4)}}\int
      \left[ -\star (e^{a}\wedge e^{b}) \wedge R_{ab}
      +\tfrac{1}{2} F\wedge \star F \right]
     \equiv
      \int \mathbf{L}\,.
\end{equation}

Notice that this theory does not contain any fields charged with respect to
the Maxwell field that can source it. The action is not invariant under any
global transformations of the Maxwell field only, but the equations of motion
supplemented with the Bianchi identity are invariant under an SO$(2)$ group of
electric-magnetic duality rotations and we are going to show in
Section~\ref{sec-electricmagneticandgeneralizedsymmetry} that they may be
consistently used in the generalized symmetric ansatz. This theory is too
simple for this ansatz to give rise to boson star solutions but we can
consider it as a ``proof of concept'' that opens the door to the use in more
complicated theories such as those we are going to in
Section~\ref{sec-EMstheories}.

As we have explained in the introduction,\footnote{See
  Ref.~\cite{Elgood:2020svt} for details on this particular case.} the
stationarity condition of the Maxwell field has to be implemented in a
gauge-invariant form, combining the standard Lie derivative with a gauge
transformation with parameter $\chi_{k}=\imath_{k}A-P_{k}$ where the
\textit{momentum map} $P_{k}$ satisfies the gauge-invariant \textit{momentum
  map equation}

\begin{equation}
  \imath_{k}F+dP_{k}
  =
  0\,,
\end{equation}

\noindent
whose integrability condition is the symmetry condition

\begin{equation}
  \delta_{k}F
  =
  -\pounds_{k}F
  =
  -\imath_{k}F
  =
  0\,.
\end{equation}

\noindent
The gauge-invariant form of stationarity condition of the Maxwell field is
the \textit{momentum map equation} 

\begin{equation}
  \label{eq:momentummapequation}
  \delta_{k}A
  =
  -\left(\pounds_{k}-\delta_{\chi_{k}}\right)A
  =
  -\left( \imath_{k}F+dP_{k}\right)
  =
  0\,.
\end{equation}

Observe that, if $k$ is timelike, $\imath_{k}F$ is the electric field for an
observer related to the time defined by $k$ and, therefore, $P_{k}$ is the
associated electrostatic potential $\Phi$ defined, as usual, up to an additive
constant. In asymptotically-flat spacetimes $\Phi$ takes a constant value at
infinity $\Phi_{\infty}$ that is purely conventional. 

In this theory it is possible to find a generic form of $\omega_{k}$ and the
generalized Komar charge of this theory is
\cite{Elgood:2020svt,Mitsios:2021zrn,Ortin:2022uxa}

\begin{equation}
  \label{eq:generalizedKomarchargeEMtheory}
  \mathbf{K}[k]
  =
  -\frac{1}{16\pi G_{N}{}^{(4)}}\star (e^{a}\wedge e^{b})P_{k\, ab}
  +\frac{1}{32\pi G_{N}{}^{(4)}}\left[P_{k}\star F -\tilde{P}_{k}F\right]\,,
\end{equation}

\noindent
where we have introduced the dual momentum map $\tilde{P}_{k}$, defined by the
\textit{dual momentum map equation}

\begin{equation}
  \label{eq:dualmomentummapequation}
  \imath_{k}\star F+d\tilde{P}_{k}
  =
  0\,,
\end{equation}

\noindent
whose local existence is guaranteed on-shell by the stationarity condition.
It can also be understood as the magnetostatic potential $\tilde{\Phi}$ and,
in asymptotically-flat spacetimes it takes an arbitrary constant value at
infinity $\tilde{\Phi}_{\infty}$.

Apart from the generalized Komar charge
Eq.~(\ref{eq:generalizedKomarchargeEMtheory}) the EM theory has another two
interesting 2-form charges: 

\begin{equation}
  \label{eq:electriccharge2formdef}
  \mathbf{Q}
  =
  \frac{1}{16\pi G_{N}^{(4)}}\star F\,,
\end{equation}

\noindent
which is closed on-shell and whose integral over a closed 2-surface gives the
electric charge $q$ enclosed by it, and

\begin{equation}
  \label{eq:magneticcharge2formdef}
  \mathbf{P}
  =
  \frac{1}{16\pi G_{N}^{(4)}} F\,,
\end{equation}

\noindent
which is closed off-shell and whose integral over a closed 2-surface gives the
magnetic charge $p$ enclosed by it. Notice that, Stokes theorem implies that
$p$ would vanish identically if $F$ was $dA$ globally.

Since the theory does not contain sources of the Maxwell field, we expect any
non-trivial stationary, asymptotically flat\footnote{Asymptotic flatness
  excludes homogeneous gravitational/electromagnetic waves which can be
  stationary, regular, horizonless and topologically trivial.} electromagnetic
fields to be sourced by singularities or sustained by non-trivial
topology. Indeed, the usual arguments applied to $\mathbf{Q}$ and $\mathbf{P}$
in boson stars lead to $q=p=0$.

Let us now turn our attention to the generalized Komar charge which can be
rewritten in the form

\begin{equation}
  \label{eq:generalizedKomarchargeEMtheory2}
  \mathbf{K}[k]
  =
  -\frac{1}{16\pi G_{N}^{(4)}}\star (e^{a}\wedge e^{b})P_{k\, ab}
  +\tfrac{1}{2}\left[\Phi\mathbf{Q} -\tilde{\Phi}\mathbf{P}\right]\,.
\end{equation}

Its integral over the 2-sphere at spatial infinity gives

\begin{equation}
  M+\Phi_{\infty}q-\tilde{\Phi}_{\infty}p
  =
  0\,,
\end{equation}

\noindent
which reduces to $M=0$ once the vanishing of the total electric and magnetic
charges has been taken into account. Again, solutions of the kind we are
looking for do not exist in the EM theory. Observe that the vanishing of $q$
and $p$ is crucial for the consistency of the result because the values of
$\Phi_{\infty}$ and $\tilde{\Phi}_{\infty}$ can be changed arbitrarily keeping
the metric (and $M$) unchanged.

It is interesting to see how the presence of a non-degenerate horizon modifies
this conclusion.

First of all, we must replace $\partial_{t}$ by the Killing vector
$k=\partial_{t}-\Omega_{H}\partial_{\varphi}$ that becomes null on the
horizon. The momentum maps $P_{k}$ and $\tilde{P}_{k}$ cannot be interpreted
as purely electrostatic and magnetostatic potentials, although we will keep
the notation $\Phi,\tilde{\Phi}$. Moreover, they do not tend to just a
constant value at infinity but their $r\rightarrow \infty$ limits may contain,
for instance, terms proportional to $\Omega_{H}\cos{\theta}$, in addition to
the arbitrary constants $\Phi_{\infty},\tilde{\Phi}_{\infty}$. Fortunately, it
can be proved that those terms do not contribute to the integrals of
$\Phi\mathbf{Q}$ and $\tilde{\Phi}\mathbf{P}$ at infinity
\cite{kn:Zattithesis} and the integral at infinity of the generalized Komar
charge Eq.~(\ref{eq:generalizedKomarchargeEMtheory2}) gives

\begin{equation}
\tfrac{1}{2}\left(M+\Phi_{\infty}q-\tilde{\Phi}_{\infty}p\right) -\Omega_{H}J\,,
\end{equation}

\noindent
while  the integral on the bifurcation surface gives

\begin{equation}
TS+\tfrac{1}{2}\left(\Phi_{H}q-\tilde{\Phi}_{H}p\right)\,,
\end{equation}

\noindent
where $\Phi_{H}$ and $\tilde{\Phi}_{H}$ are the values of the electrostatic
and magnetostatic in $\mathcal{BH}$ which are constant by virtue of the
\textit{restricted generalized zeroth law} \cite{Elgood:2020svt} which we
review in Appendix~\ref{sec-generalizedzerothlaw}.

Combining these results we obtain the Smarr formula for stationary,
asymptotically flat black holes of the EM theory

\begin{equation}
  \label{eq:SmarrformulaEMtheory}
  M = 2ST +2\Omega_{H}J
  +\left(\Phi_{H}-\Phi_{\infty}\right)q
  -\left(\tilde{\Phi}_{H}-\tilde{\Phi}_{\infty}\right)p\,.
\end{equation}

Notice that the non-vanishing electric and magnetic charges multiply
unambiguous differences of potentials.

\subsection{Generalized symmetric ansatz}
\label{sec-electricmagneticandgeneralizedsymmetry}

As we have mentioned before, the (left-hand side of the) Einstein equations
$\mathbf{E}_{a}$, Maxwell equations $\mathbf{E}$ and Bianchi identities
$\mathbf{B}$, given by\footnote{We ignore the overall factors of
  $(16\pi G_{N}^{(4)})^{-1}$ in order to simplify the expressions.}

\begin{subequations}
  \begin{align}
    \mathbf{E}_{a}
    & =
      \imath_{a}\star(e^{b}\wedge e^{c})\wedge R_{bc}
      +\tfrac{1}{2}\left(\imath_{a}F\wedge \tilde{F}
      -F\wedge \imath_{a}\tilde{F}\right)\,,
    \\
    & \nonumber \\
    \mathbf{E}
    & =
      -d \tilde{F}\,,
    \\
    & \nonumber \\
    \mathbf{B}
    & =
      -dF\,,
  \end{align}
\end{subequations}

\noindent
where we have defined the dual 2-form field strength

\begin{equation}
  \label{eq:dualfieldstrengthdef}
  \tilde{F}
  \equiv
  \star F\,.
\end{equation}

These transformations act on $F$ ad $\tilde{F}$ as

\begin{equation}
  \left(
    \begin{array}{c}
      F \\ \tilde{F} \\
    \end{array}
  \right)'
  =
    \left(
    \begin{array}{cc}
      \cos{\alpha} & \sin{\alpha} \\
      -\sin{\alpha} & \cos{\alpha} \\
    \end{array}
  \right)
  \left(
    \begin{array}{c}
      F \\ \tilde{F} \\
    \end{array}
  \right)\,,
  \hspace{1cm}
  \delta_{\alpha}
  \left(
    \begin{array}{c}
      F \\ \tilde{F} \\
    \end{array}
  \right)
  =
    \left(
    \begin{array}{cc}
     0      & \alpha \\
    -\alpha & 0 \\
    \end{array}
  \right)
  \left(
    \begin{array}{c}
      F \\ \tilde{F} \\
    \end{array}
  \right)\,.  
\end{equation}

Thus, we may implement the stationarity ansatz on these fields in a
non-trivial way, using this global symmetry, as follows:

\begin{equation}
  \label{eq:GSAinEM}
  \pounds_{k}
  \left(
    \begin{array}{c}
      F \\ \tilde{F} \\
    \end{array}
  \right)
  =
    \left(
    \begin{array}{cc}
     0      & \alpha \\
    -\alpha & 0 \\
    \end{array}
  \right)
  \left(
    \begin{array}{c}
      F \\ \tilde{F} \\
    \end{array}
  \right)\,.    
\end{equation}

Notice that this ansatz guarantees the invariance of the Maxwell
energy-momentum tensor $T_{a}$ given by 

\begin{equation}
  T_{a}
  \equiv
  \tfrac{1}{2}\left(\imath_{a}F\wedge \tilde{F}
      -F\wedge \imath_{a}\tilde{F}\right)  
\end{equation}

\noindent
under the diffeomorphism generated by the Killing vector $k$,
$\pounds_{k}T_{a}=0$.

It is convenient to introduce a dual (\textit{magnetic}) gauge field
$\tilde{A}$ such that, locally,\footnote{We are not going to address any global
  issues related to the existence of $A$ or $\tilde{A}$ in this section.}

\begin{equation}
  \tilde{F}
  \equiv
  d\tilde{A}\,.
\end{equation}

The pair $A,\tilde{A}$ transform under electric-magnetic duality as the pair
$F,\tilde{F}$, namely,

\begin{equation}
  \delta_{\alpha}
  \left(
    \begin{array}{c}
      A \\ \tilde{A} \\
    \end{array}
  \right)
  =
    \left(
    \begin{array}{cc}
     0      & \alpha \\
    -\alpha & 0 \\
    \end{array}
  \right)
  \left(
    \begin{array}{c}
      A \\ \tilde{A} \\
    \end{array}
  \right)\,.  
\end{equation}

It is worth remarking that, although this is not a symmetry of the action,
there is an on-shell conserved current associated to it:\footnote{The
  conservation is due to the well-known property
  \begin{equation}
    \star B\wedge \star A
    =
    - A\wedge B\,,
  \end{equation}
  for any pair of 2-forms $A,B$ in 4 dimensions. This property is, precisely
  the responsible for the non-invariance of the Maxwell action under
  electric-magnetic duality transformations.
}

\begin{equation}
  \label{eq:Jemdef}
  J_{em}
  \equiv
  \tfrac{1}{2}
  \left(A\wedge F +\tilde{A}\wedge \tilde{F}
  \right)\,,
  \hspace{1cm}
  dJ_{em}
  \doteq
  0\,,
\end{equation}

\noindent
which, furthermore, is invariant under electric-magnetic duality
transformations.\footnote{There is another electric-magnetic
  duality-invariant current
  \begin{equation}
      J_{em-2}
  \equiv
  \tfrac{1}{2}
  \left(A\wedge \tilde{F} -\tilde{A}\wedge F
  \right)
  =
  d\left(\tfrac{1}{2}\tilde{A}\wedge A\right)\,,
\end{equation}
but it does not seem to play any role in what follows.
}

Then, the generalized symmetric ansatz Eq.~(\ref{eq:GSAinEM}) is locally
equivalent to the conditions

\begin{equation}
d  \left(
    \begin{array}{c}
      \imath_{k}F-\alpha \tilde{A} \\ \imath_{k}\tilde{F}+\alpha A \\
    \end{array}
  \right)
  =
  0\,,
\end{equation}

\noindent
which imply the local existence of the electric and magnetic momentum maps
$P_{k},\tilde{P}_{k}$ satisfying the \textit{generalized momentum maps
  equations}

\begin{equation}
  \label{eq:generalizedmomentummapequations}
\left(
    \begin{array}{c}
      \imath_{k}F \\ \imath_{k}\tilde{F} \\
    \end{array}
  \right)
  =
  -D
  \left(
    \begin{array}{c}
      P_{k} \\ \tilde{P}_{k} \\
    \end{array}
  \right)\,,
\end{equation}

\noindent
where we have defined the covariant derivatives

\begin{equation}
    D  \left(
    \begin{array}{c}
      P_{k} \\ \tilde{P}_{k} \\
    \end{array}
  \right)
  \equiv
    d  \left(
    \begin{array}{c}
      P_{k} \\ \tilde{P}_{k} \\
    \end{array}
  \right)
  -
  \left(
    \begin{array}{cc}
     0      & \alpha \\
    -\alpha & 0 \\
    \end{array}
  \right)
  \left(
    \begin{array}{c}
      A \\ \tilde{A} \\
    \end{array}
  \right)\,,    
\end{equation}

\noindent
which are invariant under the gauge transformations

\begin{equation}
  \delta_{\chi}
  \left(
    \begin{array}{c}
      A \\ \tilde{A} \\
    \end{array}
  \right)
  =
  d  \left(
    \begin{array}{c}
      \chi \\ \tilde{\chi} \\
    \end{array}
  \right)
\,,    
\hspace{1cm}
  \delta_{\chi}
  \left(
    \begin{array}{c}
      P_{k} \\ \tilde{P}_{k} \\
    \end{array}
  \right)
  =
  \left(
    \begin{array}{cc}
     0      & \alpha \\
    -\alpha & 0 \\
    \end{array}
  \right)
  \left(
    \begin{array}{c}
      \chi \\ \tilde{\chi} \\
    \end{array}
  \right)\,.      
\end{equation}

The momentum maps transform as St\"uckelberg fields and we may eliminate them
using the appropriate gauge transformation, if necessary. Thus, it is clear
that they cannot be identified with electrostatic or magnetostatic
potentials as in the symmetric case. 

To check the consistency of this construction, it is interesting to review it
in terms of the gauge fields.  Taking into account their gauge freedoms, the
generalized symmetric ansatz reads

\begin{equation}
  \label{eq:GSAinEMforgaugefields}
  \pounds_{k}
    \left(
    \begin{array}{c}
      A \\ \tilde{A} \\
    \end{array}
  \right)
  =
  d
  \left(
    \begin{array}{c}
      \chi_{k} \\ \tilde{\chi}_{k} \\
    \end{array}
  \right)
  +    \left(
    \begin{array}{cc}
     0      & \alpha \\
    -\alpha & 0 \\
    \end{array}
  \right)
  \left(
    \begin{array}{c}
      A \\ \tilde{A} \\
    \end{array}
  \right)\,,    
\end{equation}

\noindent
and we can immediately see that it is satisfied by the choice of parameters of
the compensating gauge transformations

\begin{equation}
  \left(
    \begin{array}{c}
      \chi_{k} \\ \tilde{\chi}_{k} \\
    \end{array}
  \right)
  =
    \left(
    \begin{array}{c}
      \imath_{k}A-P_{k} \\ \imath_{k}\tilde{A}-\tilde{P}_{k} \\
    \end{array}
  \right)\,.
\end{equation}

In black-hole spacetimes, if $k=\partial_{t}-\Omega_{H}\partial_{\varphi}$ is
the Killing vector that characterizes the event horizon as a Killing horizon,
in the definition of the generalized symmetric ansatz we have to use the same
global symmetry for $\partial_{t}$ (with parameter $\omega$) and
$\partial_{\varphi}$ (with parameter $m$) and the covariant derivatives of the
momentum maps are now

\begin{equation}
    D  \left(
    \begin{array}{c}
      P_{k} \\ \tilde{P}_{k} \\
    \end{array}
  \right)
  \equiv
    d  \left(
    \begin{array}{c}
      P_{k} \\ \tilde{P}_{k} \\
    \end{array}
  \right)
  -
(\omega-\Omega_{H}m)  \left(
    \begin{array}{cc}
     0      & 1 \\
    -1 & 0 \\
    \end{array}
  \right)
  \left(
    \begin{array}{c}
      A \\ \tilde{A} \\
    \end{array}
  \right)\,.    
\end{equation}

On the horizon, both $\imath_{k}F$ and $\imath_{k}\tilde{F}$ vanish
identically. In the symmetric case the generalized zeroth law can be derived
from this fact (see Appendix~\ref{sec-generalizedzerothlaw}). In this case, we
find that 

\begin{equation}
    D  \left(
    \begin{array}{c}
      P_{k} \\ \tilde{P}_{k} \\
    \end{array}
  \right)
  \stackrel{\mathcal{H}}{=}
  0\,.
\end{equation}

If the synchronization condition Eq.~(\ref{eq:synchronizationcondition}) is
not satisfied $\omega-\Omega_{H}m\neq 0$, these equations imply that both
connections $A$ and $\tilde{A}$ are pure gauge on the horizon
and, thus, the field strength and its dual vanish identically there

\begin{equation}
      \left(
    \begin{array}{c}
      F \\ \tilde{F} \\
    \end{array}
  \right)
  \stackrel{\mathcal{H}}{=}
  0\,.
\end{equation}

The generalized symmetric ansatz Eq.~(\ref{eq:GSAinEM}) does not modify the
definitions of electric and magnetic charges satisfying Gauss laws
$\mathbf{Q},\mathbf{P}$ Eqs.~(\ref{eq:electriccharge2formdef}) and
(\ref{eq:magneticcharge2formdef}).  Since we can compute the electric and
magnetic charges integrating these field strengths over any section of the
horizon obtaining the same result (because it is the same result one obtains
integrating over the 2-sphere at spatial infinity), we conclude that, in this
case, the generalized symmetric ansatz leads to vanishing electric and
magnetic charges $q=p=0$ in black-hole spacetimes. In boson-star spacetimes
they vanish for the same reasons as in the previous cases.

If the synchronization condition Eq.~(\ref{eq:synchronizationcondition}) is
satisfied, then the momentum maps admit the standard interpretation of
electrostatic and magnetostatic potentials and satisfy the generalized zeroth
law and the electric and magnetic charges need not vanish.

The generalized Komar charge is modified. The Noether--Wald charge
$\mathbf{Q}[k]$ only depends on diffeomorphisms and the induced gauge
transformations and it is not modified and it is still given by
\cite{Elgood:2020svt}

\begin{equation}
  \mathbf{Q}[\xi]
  =
  \star (e^{a}\wedge e^{b})P_{\xi\, ab} -P_{\xi}\tilde{F}\,.
\end{equation}

\noindent
On-shell, it satisfies

\begin{equation}
  d\mathbf{Q}[\xi]
  \doteq
  \mathbf{\Theta}(e,A,\delta_{\xi}e,\delta_{\xi}A)
  +\imath_{\xi}\mathbf{L}\,,
\end{equation}

\noindent
where the presymplectic potential and the on-shell Lagrangian are given by 

\begin{subequations}
  \begin{align}
  \mathbf{\Theta}(e,A,\delta_{\xi}e,\delta_{\xi}A)
    & =
        -\star(e^{a}\wedge e^{b})\wedge \delta_{\xi}\omega_{ab}
    -\tilde{F}\wedge \left(\imath_{\xi}F+dP_{\xi}\right)\,,
    \\
    & \nonumber \\
    \mathbf{L}
    & \doteq
        \tfrac{1}{2} F\wedge \tilde{F}\,.
  \end{align}
\end{subequations}

Now, when $\xi=k$, the first term in $\mathbf{\Theta}$ vanishes but the second
does not, according to the generalized momentum map
equations~(\ref{eq:generalizedmomentummapequations}), which we also have to
apply to the calculation of $\imath_{k}\mathbf{L}$. Substituting the result in
$d\mathbf{Q}[k]$, we get

\begin{equation}
d\left[-\star (e^{a}\wedge e^{b})P_{k\, ab}
  +\tfrac{1}{2}\left[P_{k}\tilde{F} -\tilde{P}_{k}F\right]\right]
\doteq
-\omega J_{em}\,.
\end{equation}

\noindent
where $J_{em}$ has been defined in Eq.~(\ref{eq:Jemdef}).  Since $J_{em}$ is
closed on-shell, there must be a 2-form $\mathcal{J}_{em}$, whose form depends
on the particular solution on which we evaluate it, such that

\begin{equation}
  d \mathcal{J}_{em}
  \doteq
  J_{em}\,,
\end{equation}

\noindent
and we arrive at the following generalized Komar charge

\begin{equation}
  \mathbf{K}[k]
  =
  -\star (e^{a}\wedge e^{b})P_{k\, ab}
  +\tfrac{1}{2}\left[P_{k}\tilde{F} -\tilde{P}_{k}F\right]
  +\omega \mathcal{J}_{em}\,.
\end{equation}

Since the electric and magnetic charges must vanish for any hypothetical boson
star, it seems unlikely that the integral of $\mathcal{J}_{em}$ at infinity
could help us to avoid the conclusion $M=0$. In Appendix~\ref{sec-trial} we
have studied, as a proof of concept, a simple example of time-dependent
solutions of the Maxwell equations in a (non-back-reacted) stationary
spacetime (Minkowski spacetime), showing their existence. They turn out to be
superpositions of electromagnetic waves with no electric nor magnetic
charges.

In black hole spacetimes the last term must be replaced by
$(\omega-\Omega_{H}m)\mathcal{J}_{em}$ and there are two two possible cases:
when the synchronization condition is not satisfied we must use the vanishing
of $F$ and $\tilde{F}$ on the bifurcation surface because the momentum maps do
not satisfy a restricted generalized zeroth law and, again, the vanishing of
the electric and magnetic charges makes it very unlikely that the current
$\mathcal{J}_{em}$ gives any finite contribution at infinity. When the
synchronization condition is satisfied, the last term vanishes identically and
the momentum maps are constant over the horizon and we recover exactly the
same Smarr formula as in the symmetric case. It remains to be seen if there
are any black hole solutions satisfying this ansatz but, in principle, the
Smarr formula does not exclude this possibility.

\section{The Einstein--scalar theory with a scalar potential}
\label{sec-Estheoryscalarpotential}

The thermodynamics of the black holes of this theory was recently studied by
us in Ref.~\cite{Ballesteros:2023muf}. Thus, we shall be brief. The action is
the one considered in Eq.~(\ref{eq:Esaction}) plus a scalar potential term:

\begin{equation}
  \label{eq:Espluspotentialaction}
    S[e,\phi]
    =
    \frac{1}{16\pi G_{N}^{(4)}} \int \left\{
      -\star (e^{a}\wedge e^{b})\wedge R_{ab}
      +\tfrac{1}{2}d\phi\wedge \star d\phi +\star V(\phi)
      \right\}
      \equiv
      \int \mathbf{L}\,.
\end{equation}

The scalar potential $V(\phi)$ is assumed not to be constant, so it cannot be
interpreted as a cosmological constant. Therefore, there is no shift symmetry,
no associated on-shell conserved Noether current and no possible generalized
symmetric ansatz in this theory. In spite of this, an on-shell-closed 2-form
charge can be defined in stationary solutions using the following observation
\cite{Pacilio:2018gom,Ballesteros:2023iqb}: if all the fields of the solution
we are considering are invariant under the diffeomorphism generated by $k$, we
have, on the one hand

\begin{equation}
    -\imath_{k}d\star d\phi
    =
    \left(d\imath_{k} -\pounds_{k}\right)\star d\phi
    =
      d\imath_{k}\star d\phi\,,
\end{equation}

\noindent
and on the other hand

\begin{equation}
  0
  =
  \pounds_{k}\star V'
  =
  d\imath_{k}\star V'\,,
  \hspace{1cm}
  V'
  \equiv
  \frac{\partial V}{\partial\phi}\,,
\end{equation}

\noindent
which implies the local existence of a 2-form $\mathcal{W}_{k}$ such that

\begin{equation}
    \label{eq:Wkdefinition}
  \imath_{k}\star V'
  \doteq
  d\mathcal{W}_{k}\,.
\end{equation}

In this case it is not possible to find a generic expression for
$\mathcal{W}_{k}$. Its form will depend on the particular solution on which
$\star V'$ is evaluated but we can always add a closed 2-form to it so that,
for asymptotically flat solutions $\mathcal{W}_{k}(\infty)=0$.

Then, if we take the interior product of $k$ with the scalar equation of
motion and use the above results

\begin{equation}
  \begin{aligned}
  \imath_{k}\mathbf{E}_{\phi}
  & =
\frac{1}{16\pi G_{N}^{(4)}}\imath_{k}\left[-d\star d\phi+\star V'\right]
    \\
    & \\
    & =
      d\left\{\frac{1}{16\pi G_{N}^{(4)}}
      \left[\imath_{k}\star d\phi+\mathcal{W}_{k}\right]\right\}\,,
  \end{aligned}
\end{equation}

\noindent
and we can define the on-shell closed 2-form charge\footnote{The normalization
  is purely conventional.}

\begin{equation}
  \mathbf{Q}_{\phi}[k]
  \equiv
  -\frac{1}{4\pi G_{N}^{(4)}}
  \left[\imath_{k}\star d\phi+\mathcal{W}_{k}\right]\,,  
\end{equation}

\noindent
whose integral over closed 2-dimensional surfaces gives, by definition, the
scalar charge $\Sigma$ enclosed in it. At infinity, because of the boundary
condition $\mathcal{W}_{k}(\infty)=0$, the scalar charge depends only on the
first term in the above equation. Furthermore, the usual argument leads to
$\Sigma=0$. In presence of a non-degenerate horizon we find that $\Sigma$ is
given by the integral of $\mathcal{W}_{k}$ on the bifurcation sphere. Thus,
$\Sigma$ would be secondary hair, dependent on the dimensionful constants
defining $V$.

The generalized Komar charge is in this theory \cite{Ballesteros:2023iqb}

\begin{equation}
  \label{eq:EspluspotentialgeneralizedKomarcharge}
  \mathbf{K}[k]
  =
  -\frac{1}{16\pi G_{N}^{(4)}}\left[\star (e^{a}\wedge e^{b})P_{k\, ab}
    -\mathcal{V}_{k} \right]\,,
\end{equation}

\noindent
where, in the same vein as $\mathcal{W}_{k}$, the 2-form $\mathcal{V}_{k}$ is
defined to satisfy

\begin{equation}
  \label{eq:Vkdef}
  d\mathcal{V}_{k}
  \doteq
  -\imath_{k}\star V\,.
\end{equation}

\noindent
Its value at spatial infinity can be set to zero for asymptotically-flat
solutions, $\mathcal{V}_{k}(\infty)=0$, upon the addition of a closed 2-form.

With the chosen boundary conditions, the integral of $\mathbf{K}[k]$ at
spatial infinity gives, yet again, $M=0$. When the potential is not
definite-positive, this may not be enough to ensure that the only possible
solution is Minkowski spacetime because the positive mass theorem would not be
valid.

Allowing for a non-degenerate horizon and choosing $k$ accordingly, we get the
Smarr formula \cite{Ballesteros:2023iqb}

\begin{equation}
  \label{eq:Smarrformula2}
    M
     =
     2ST +2\Omega_{H}J +2\alpha \Phi_{\alpha}\,,
     \,\,\,\,\,
     \text{where}
     \,\,\,\,\,
     \Phi_{\alpha}
     \equiv
     -\frac{1}{16\pi G_{N}^{(4)}}
  \int_{\Sigma^{3}}\imath_{k}\star \frac{\partial V}{\partial \alpha}\,,
\end{equation}

\noindent
where $\alpha$ is a dimensionful constant in $V$ which plays the role of a new
thermodynamic variable while $\Phi_{\alpha}$ plays the role of its conjugate
potential \cite{Hajian:2021hje,Ortin:2021ade,Meessen:2022hcg}. If the scalar
potential depends on more dimensionful constants, $\alpha^{i}$, the last term
is replace by a sum over similar terms.

\section{The Einstein--Maxwell--scalar theories}
\label{sec-EMstheories}

In this section we are going to consider a generic supergravity-inspired
theory that includes the three cases we have studied so far as particular
examples. This theory contains, apart from the gravitational field $e^{a}$,
$n_{v}$ Abelian gauge fields $A^{\Lambda}$, $\Lambda=1,\ldots,n_{V}$. with
field strengths $F^{\Lambda}\equiv dA^{\Lambda}$ and $n_{S}$ scalar fields
$\phi^{x}$, $x=1,\ldots,n_{s}$ that parametrize a non-linear $\sigma$-model
with positive-definite target-space metric $g_{xy}(\phi)$ coupling to
themselves via a scalar potential $V(\phi)$ and to the Abelian gauge fields
via the symmetric, scalar-dependent, matrices $I_{\Lambda\Sigma}(\phi)$ (which
is conventionally assumed to be negative-definite) and
$R_{\Lambda\Sigma}(\phi)$. The action takes the form\footnote{In supergravity
  theories, the numbers $n_{V}$ and $n_{s}$, the $\sigma$-model metric, the
  scalar matrices and the scalar potential are constrained by
  supersymmetry. In supergravity, the scalar potential would be associated to
  the gauging of R-symmetry. The gauging of other symmetries would give rise
  to couplings not concluded in the action Eq.~(\ref{eq:EMsaction}). In
  particular, the gauging of symmetries which act as isometries of the
  $\sigma$-model metric leads to the replacement of the derivatives of the
  scalar fields by covariant derivatives. In some cases the gauge symmetries
  can be used to completely gauge away some of the scalars, whose kinetic
  terms become mass terms for the vector fields. The so-called
  Einstein--Proca--Higgs model considered in
  Refs.~\cite{Herdeiro:2023lze,Herdeiro:2024pmv} can be obtained in this way
  starting from a model of the form Eq.~(\ref{eq:EMsaction}). We will study
  this more general class of models elsewhere \cite{kn:BFMOP}.}

\begin{equation}
\label{eq:EMsaction}
\begin{aligned}
  S
  & =
\frac{1}{16\pi G_{N}^{(4)}}\int 
  \left[ -\star (e^{a}\wedge e^{b}) \wedge R_{ab}
    +\tfrac{1}{2}g_{xy}d\phi^{x}\wedge \star d\phi^{y}
  \right.
  \\
  & \\
  & \hspace{.5cm}
  \left.
    -\tfrac{1}{2}I_{\Lambda\Sigma}F^{\Lambda}\wedge \star F^{\Sigma}
    -\tfrac{1}{2}R_{\Lambda\Sigma}F^{\Lambda}\wedge F^{\Sigma}
+\star V\right]\,,
\end{aligned}
\end{equation}

\noindent
The equations of motion $\mathbf{E}_{a},\mathbf{E}_{x},\mathbf{E}_{\Lambda}$
and the presymplectic potential $\mathbf{\Theta}(\varphi,\delta\varphi)$ (here
$\varphi$ stands for all the fields of the theory) are defined by the general
variation of the fields in the action

\begin{equation}
  \delta S
  =
  \int\left\{
    \mathbf{E}_{a}\wedge \delta e^{a} + \mathbf{E}_{x}\delta\phi^{x}
    +\mathbf{E}_{\Lambda}\delta A^{\Lambda}  
    +d\mathbf{\Theta}(\varphi,\delta\varphi)
    \right\}\,.
\end{equation}

\noindent
Ignoring the overall factors of $(16\pi G_{N}^{(4)})^{-1}$, they are given by

\begin{subequations}
  \begin{align}
    \label{eq:Ea}
    \mathbf{E}_{a}
    & =
      \imath_{a}\star(e^{b}\wedge e^{c})\wedge R_{bc}
      +\tfrac{1}{2}g_{xy}\left(\imath_{a}d\phi^{x} \star d\phi^{y}
      +d\phi^{x}\wedge \imath_{a}\star d\phi^{y}\right)
      \nonumber \\
    & \nonumber \\
    & \hspace{.5cm}
      -\tfrac{1}{2}I_{\Lambda\Sigma}\left(\imath_{a}F^{\Lambda}\wedge\star
      F^{\Sigma} -F^{\Lambda}\wedge \imath_{a}\star F^{\Sigma}\right)
      -\imath_{a}\star V\,,
    \\
    & \nonumber \\
    \mathbf{E}_{x}
    & =
      -g_{xy}\left\{d\star d\phi^{y}
      +\Gamma_{zw}{}^{y}d\phi^{z}\wedge\star d\phi^{w} \right\}
      \nonumber \\
    & \nonumber \\
    & \hspace{.5cm}
      -\tfrac{1}{2}\partial_{x}I_{\Lambda\Sigma} F^{\Lambda}\wedge\star
      F^{\Sigma}
      -\tfrac{1}{2}\partial_{x}R_{\Lambda\Sigma} F^{\Lambda}\wedge F^{\Sigma}
       +\star \partial_{x}V\,,
    \\
    & \nonumber \\
    \mathbf{E}_{\Lambda}
    & =
      d F_{\Lambda}\,,
    \\
    & \nonumber \\
      \label{eq:Theta}
    \mathbf{\Theta}(\varphi,\delta\varphi)
    & =
      -\star (e^{a}\wedge e^{b})\wedge \delta \omega_{ab}
      +g_{xy}\star d\phi^{x}\delta\phi^{y}
      -F_{\Lambda}\wedge \delta A^{\Lambda}\,.
  \end{align}
\end{subequations}

\noindent
where we have defined the dual 2-form field strength

\begin{equation}
  \label{eq:dualfieldstrengthsdef}
  F_{\Lambda}
  \equiv
  I_{\Lambda\Sigma}\star F^{\Sigma}+R_{\Lambda\Sigma}F^{\Sigma}\,.
\end{equation}

\subsection{Global symmetries}
\label{sec-globalsymmetries}

In supergravity theories the symmetries of the $\sigma$-model, generated by
the Killing vectors $K_{I}{}^{x}(\phi)$ of $g_{xy}(\phi)$

\begin{equation}
  \label{eq:gxyinvariance}
  \delta_{I}g_{xy}
  =
  -\pounds_{K_{I}}g_{xy}
  =
  0\,,
\end{equation}

\noindent
and the symmetries of the Abelian gauge fields are related\footnote{See,
  \textit{e.g.}~Section~1 of Ref.~\cite{Ballesteros:2023iqb} for a detailed
  description of the properties of the scalar matrices that are required for
  this relation to work.} thanks to the equivariance property

\begin{subequations}
  \begin{align}
 \label{eq:equivarianceconditionR}
    \delta_{I}R_{\Lambda\Sigma}
    & =
  T_{I\, \Lambda\Sigma}
  +T_{I}{}_{\Lambda}{}^{\Omega}R_{\Omega\Sigma}
  -R_{\Lambda\Omega}T_{I}{}^{\Omega}{}_{\Sigma}
  -R_{\Lambda\Gamma} T_{I}{}^{\Gamma\Omega}R_{\Omega\Sigma}
  +I_{\Lambda\Gamma} T_{I}{}^{\Gamma\Omega}I_{\Omega\Sigma}\,,
    \\
    & \nonumber \\
 \label{eq:equivarianceconditionI}
    \delta_{I}I_{\Lambda\Sigma}
    & =
  T_{I}{}_{\Lambda}{}^{\Omega}I_{\Omega\Sigma}
  -I_{\Lambda\Omega}T_{I}{}^{\Omega}{}_{\Sigma}
  -2R_{(\Lambda|\Gamma} T_{I}{}^{\Gamma\Omega}I_{\Omega|\Sigma)}\,.
  \end{align}
\end{subequations}

\noindent
We will assume that this is the case here. Otherwise we would restrict
ourselves to the subgroup of isometries of the $\sigma$-model metric for which
the above equivariance conditions hold. Thus, in absence of the scalar
potential, which may break some of those symmetries, the equations of motion
are invariant under the following transformations of the scalars and gauge
fields

\begin{subequations}
  \begin{align}
    \delta_{I}\phi^{x}
    & =
      K_{I}{}^{x}(\phi)\,,
    \\
    & \nonumber \\
    \label{eq:deltaIFM}
    \delta_{I} F^{M}
    & =
      T_{I}{}^{M}{}_{N}F^{N}\,,
  \end{align}
\end{subequations}

\noindent
where we have defined the symplectic vectors of field strengths

\begin{equation}
  \left(F^{M}\right)
  \equiv
  \left(
    \begin{array}{c}
      F^{\Lambda} \\ F_{\Lambda} \\
    \end{array}
    \right)\,,
\end{equation}

\noindent
and where the matrices $T_{I}$ in Eqs.~(\ref{eq:equivarianceconditionR}) are
blocks of the matrices 
$T_{I}{}^{M}{}_{N}$ in Eq.~(\ref{eq:deltaIFM}), \textit{i.e.}

\begin{equation}
  \left(T_{I}{}^{M}{}_{N} \right)
  =
  \left(
    \begin{array}{cc}
      T^{\Lambda}{}_{\Sigma}  & T^{\Lambda\Sigma} \\
      T_{\Lambda\Sigma} & T_{\Lambda}{}^{\Sigma} \\
    \end{array}
  \right)\,.
\end{equation}

The Killing vectors $K_{I}$ and the $2n_{V}\times 2n_{V}$ matrices
$T_{I}$ satisfy the Lie algebra

\begin{subequations}
  \begin{align}
    [K_{I},K_{J}]
    & =
      -f_{IJ}{}^{K}K_{K}\,,
    \\
    & \nonumber \\
    [T_{I},T_{J}]
    & =
      +f_{IJ}{}^{K}T_{K}\,.
  \end{align}
\end{subequations}

\noindent
Furthermore, the matrices $T_{I}$ belong to the Lie algebra of the symplectic
group SL$(2n_{V},\mathbb{R})$ in the fundamental (vector) representation
\cite{Gaillard:1981rj}, which implies for
the block matrices

\begin{equation}
  \begin{aligned}
    T_{A\, \Lambda\Sigma}
    & = 
    T_{A\, \Sigma\Lambda}\,,
    \\
    & \\
    T_{A}{}^{\Lambda}{}_{\Sigma}
    & =
    -T_{A}{}_{\Sigma}{}^{\Lambda}\,,
    \\
    & \\
    T_{A}{}^{\Lambda\Sigma}
    & = 
    T_{A}{}^{\Sigma\Lambda}\,.    
  \end{aligned}
\end{equation}

Some of the transformations of the field strengths are electric-magnetic
duality transformations transforming $F^{\Lambda}$ into $F_{\Lambda}$

\begin{equation}
    \delta_{I} F^{\Lambda}
    =
    T_{I}{}^{\Lambda}{}_{\Sigma}F^{\Sigma}
    +T_{I}{}^{\Lambda\Sigma}F_{\Sigma}\,.
\end{equation}

\noindent
Those transformations are symmetries of the equations of motion but they do
not leave the action invariant. As we have seen in
Section~\ref{sec-electricmagneticandgeneralizedsymmetry} they can also be used
to formulate a generalized symmetric ansatz.

In presence of an scalar potential, we are going to assume that a subset of
these symmetries whose generators we will label with $A,B,C,\ldots$
($K_{A},T_{A}$) generate a Lie subalgebra with structure constants
$f_{AB}{}^{C}$ are preserved. In particular, we assume that

\begin{equation}
  \label{eq:Vinvariance}
  \pounds_{K_{A}}V
  =
  K_{A}{}^{x}\partial_{x}V
  =
  0\,.
\end{equation}

We are going to denote with indices $U,V,W,\dots$ the Killing vectors that do
not leave invariant the scalar potential

\begin{equation}
  \label{eq:Vnoninvariance}
  \pounds_{K_{U}}V
  =
  K_{U}{}^{x}\partial_{x}V
  \neq
  0\,.
\end{equation}

\subsection{Charges and Smarr formula for the symmetric ansatz}

Contracting the scalar equations of motion $\mathbf{E}_{x}$ with the Killing
vectors $K_{I}{}^{x}$ we get \cite{Bandos:2016smv}:

\begin{subequations}
  \begin{align}
  \label{eq:dJAdef}
  K_{I}{}^{x} \mathbf{E}_{x}
  & \equiv
    dJ_{I}+\star K_{I}{}^{x}\partial_{x}V
    +\tfrac{1}{2}\Omega_{MP}T_{A}{}^{P}{}_{N}A^{M}\wedge \mathbf{E}^{N}\,,
    \\
    & \nonumber \\
    \label{eq:JIdef}
  J_{I}
 & =
   -K_{I\, x}\star d\phi^{x}
   -\tfrac{1}{2}\Omega_{MP}T_{I}{}^{P}{}_{N}A^{M}\wedge F^{N}\,,
  \end{align}
\end{subequations}

\noindent
where we have used Eqs.~(\ref{eq:gxyinvariance}),
(\ref{eq:equivarianceconditionR}) and (\ref{eq:Vinvariance}) and where 

\begin{equation}
  \left(\Omega_{MN}\right)
  = 
  \left(
    \begin{array}{lr}
      0 & \mathbb{1}_{n_{V}\times n_{V}}   \\
      & \\
-\mathbb{1}_{n_{V}\times n_{V}}      & 0 \\
    \end{array}
  \right)\,,
\end{equation}

In absence of scalar potential, the 3-forms $J_{I}$ would be the
on-shell-closed \textit{Noether--Gaillard--Zumino (NGZ)} currents of the
theory \cite{Gaillard:1981rj}. The scalar potential breaks some of the global
symmetries and we end up with the on-shell-closed NGZ currents $J_{A}$ and the
currents $J_{U}$ which are not closed on-shell:

\begin{subequations}
  \begin{align}
    \label{eq:dJA0}
    dJ_{A}
    & \doteq
      0\,,
    \\
    & \nonumber \\
    dJ_{U}
    & \doteq
      -\star K_{U}{}^{x}\partial_{x}V\,.
  \end{align}
\end{subequations}

\noindent
In stationary solutions in which all the fields satisfy the symmetric ansatz
(invariance without compensating global transformations\footnote{As we have
  stressed, for the gauge fields, we always need to introduce a
  ``compensating'' gauge transformation.}), the usual symmetry arguments lead
to two kinds of on-shell-closed 2-form charges \cite{Ballesteros:2023iqb}

\begin{subequations}
  \begin{align}
    \label{eq:QAkcharges}
    \mathbf{Q}_{A}[k]
    & \equiv
      \frac{1}{4\pi G_{N}^{(4)}}\left\{\imath_{k}K_{A\, x}\star d\phi^{x}
      +\Omega_{MP}T_{A}{}^{P}{}_{N}P_{k}{}^{M}F^{N}\right\}\,,
    \\
    & \nonumber \\
    \label{eq:QUkcharges}
    \mathbf{Q}_{U}[k]
    & \doteq
      \frac{1}{4\pi G_{N}^{(4)}}\left\{      \imath_{k}K_{U\, x}\star d\phi^{x}
      +\Omega_{MP}T_{U}{}^{P}{}_{N}P_{k}{}^{M}F^{N}
      +\mathcal{W}_{U,k}\right\}\,,
  \end{align}
\end{subequations}

\noindent
where the 2-forms $\mathcal{W}_{U,k}$ are defined by

\begin{equation}
  d\mathcal{W}_{U,k}
  \equiv
  \imath_{k}\star K_{U}{}^{x}\partial_{x}V\,,
\end{equation}

\noindent
and where we have defined the symplectic vector of momentum maps

\begin{equation}
  \label{eq:momentummapequationsymplectic}
\left(  dP_{k}{}^{M}\right)
  \equiv
\left(  \imath_{k}F^{M}\right)\,.
\end{equation}

As in the case considered in Section~\ref{sec-Estheoryscalarpotential}, we
cannot give a generic expression for the 2-forms $\mathcal{W}_{U,k}$ because it
will depend on the particular solution considered.

The generalized Komar charge of this theory (symmetric ansatz) can be easily
found combining the results and strategies of the previous sections with those
of Refs.~\cite{Mitsios:2021zrn,Ortin:2022uxa}\footnote{The derivation is
  reviewed to account for the generalized symmetric ansatz in
  Section~\ref{sec-EMSgeneralizedsymmetricansatz}, anyway.} and it is given by

\begin{equation}
  \label{eq:generalizedKomarchargeEMstheory}
  \mathbf{K}[k]
  =
  -\frac{1}{16\pi G_{N}{}^{(4)}}\left\{\star (e^{a}\wedge e^{b})P_{k\, ab}
  +\tfrac{1}{2}\left[P_{k}{}^{\Lambda}F_{\Lambda}
    -P_{k\, \Lambda}F^{\Lambda}\right]
  -\mathcal{V}_{k}\right\}\,,
\end{equation}

\noindent
where $\mathcal{V}_{k}$ has been defined in Eq.~(\ref{eq:Vkdef}).

Finally, the electric $q_{\Lambda}$ and magnetic $p^{\Lambda}$ charges,
combined in a symplectic vector of charges $q^{M}$ can be defined as the
integrals of the symplectic vector of 2-form charges

\begin{equation}
  \label{eq:QMdef}
  \mathbf{Q}^{M}
  \equiv
  \frac{1}{16\pi G_{N}^{(4)}}F^{M}\,,
  \hspace{1cm}
  q^{M}
  \equiv
  \int_{S^{2}_{\infty}}\mathbf{Q}^{M}\,.
\end{equation}

As usual, with the trivial implementation of the stationarity condition,
electric, magnetic and scalar charges must vanish
$q^{M}=\Sigma_{A}=\Sigma_{U}=0$. Taking this into account, the integral of the
exterior derivative of the Komar charge over a Cauchy surface of a boson star
gives, yet again, $M=0$.

Allowing for a non-degenerate horizon and choosing $k$ accordingly, we get a
generalization of the Smarr formulas obtained before, that combines them:

\begin{equation}
  \label{eq:Smarrformula3}
  M
  =
  2ST +2\Omega_{H}J +(\Phi^{\Lambda}_{H}-\Phi^{\Lambda}_{\infty})q_{\Lambda}
  -(\Phi_{\Lambda\,  H}-\Phi_{\Lambda\, \infty})p^{\Lambda}
  +2\alpha\Phi_{\alpha}\,,
\end{equation}

\noindent
where $\alpha$ and $\Phi_{\alpha}$ have the same meaning as in
Eq.~(\ref{eq:Smarrformula2}).

\subsection{Generalized symmetric ansatz}
\label{sec-EMSgeneralizedsymmetricansatz}

The theories we are considering can have a large number of global symmetries,
with different kinds of orbits. This means that, in stationary, axisymmetric
spacetimes, it may be possible to define a generalized symmetric ansatz
involving the Killing vector $\partial_{\varphi}$ as well as
$\partial_{t}$:\footnote{See footnote~\ref{foot:esafootnote} in
  page~\pageref{foot:esafootnote}.  Notice that this is the most general thing
  we can do: as proven in Appendix~\ref{sec-consistencycondition}, if we only
  use the Killing vector $\partial_{\varphi}$ the spacetime cannot be
  spherically symmetric nor it can admit any other Killing vector whose Lie
  brackets with $\partial_{\varphi}$ do not vanish. In black-hole spacetimes
  this is usually associated to rotation.}

\begin{subequations}
  \begin{align}
    \partial_{t}\phi^{x}
    & =
      \vartheta_{t}{}^{A} K_{A}{}^{x}
      \equiv
      \omega K^{x}\,,
    \\
    & \nonumber \\
    \partial_{\varphi}\phi^{x}
    & =
      \vartheta_{\varphi}{}^{A} K_{A}{}^{x}(\phi)
      \equiv
      mL^{x}\,,
  \end{align}
\end{subequations}

\noindent
for two constants $\omega,m$ and two Killing vectors of the target space
metric $K,L$ that also leave invariant the scalar potential $V$. In the
particular case of the massive, complex, Klein--Gordon field there is only one
isometry that leaves invariant the scalar potential and, therefore, $K=L$ was
the only possibility considered in
Ref.~\cite{Herdeiro:2014goa,Herdeiro:2015gia}.

Since the global symmetries of these theories also act on the gauge field
strengths and on their duals according to Eq.~(\ref{eq:deltaIFM}) we must also
include them in the ansatz:

\begin{subequations}
  \begin{align}
    \pounds_{t}F^{M}
    & =
      \vartheta_{t}{}^{A} T_{A}{}^{M}{}_{N}F^{N}
      \equiv
      \omega T^{M}{}_{N}F^{N}\,,
    \\
    & \nonumber \\
    \pounds_{\varphi}F^{M}
    & =
      \vartheta_{\varphi}{}^{A} T_{A}{}^{M}{}_{N}F^{N}
      \equiv
      m S^{M}{}_{N}F^{N}\,.
  \end{align}
\end{subequations}

Since $\partial_{t}$ and $\partial_{\varphi}$ have vanishing spacetime Lie
brackets, the matrices $T$ and $S$ must commute and $K$ and $L$ must have
vanishing target space Lie brackets as well. Thus, we can use target space
coordinates (scalar fields) adapted to them. If we call them, respectively,
$\phi^{1}$ and $\phi^{2}$, the ansatz reads

\begin{subequations}
  \begin{align}
    \partial_{t}\phi^{x}
    & =
      \omega \delta^{x}{}_{1}\,,
    \\
    & \nonumber \\
    \partial_{\varphi}\phi^{x}
    & =
      m  \delta^{x}{}_{2}\,,
    \\
    & \nonumber \\
      \Rightarrow
      \,\,\,\,
    \phi^{x}
    &
      =
      \delta^{x}{}_{1}\omega t
      +\delta^{x}{}_{2}m\varphi
      +f^{x}(x^{1},x^{2})\,,
  \end{align}
\end{subequations}

\noindent
and, if $K=L$, then $\phi^{2}=\phi^{1}$,  $T=S$ and

\begin{subequations}
  \begin{align}
    \partial_{t}\phi^{x}
    & =
      \omega \delta^{x}{}_{1}\,,
    \\
    & \nonumber \\
    \partial_{\varphi}\phi^{x}
    & =
      m  \delta^{x}{}_{1}\,.
    \\
    & \nonumber \\
      \Rightarrow
      \,\,\,\,
    \phi^{x}
    &
      =
      \delta^{x}{}_{1}\left(\omega t
      +m\varphi \right)
      +f^{x}(x^{1},x^{2})\,,
  \end{align}
\end{subequations}

As we have mentioned before, it is usually stated that the solutions
satisfying this ansatz are neither stationary nor antisymmetric and that their
only spacetime symmetry, present when $K=L$ is the one generated by the
Killing vector

\begin{equation}
  \label{eq:kcombination}
  k
  \equiv
  \partial_{t}-\frac{\omega}{m}\partial_{\varphi}\,.
\end{equation}

As we have explained at length in the introduction, the above ansatz follows
from a modified (``twisted'') definition of what being stationary and
axisymmetric means. In other words, it is a different implementation of
stationarity and axisymmetry and we are going to see that one can define
on-shell closed Komar charges for each Killing vector $\partial_{t}$ and
$\partial_{\varphi}$, independently.  The combination $k$ in
Eq.~(\ref{eq:kcombination}) is clearly special, though, and it can be
shown\footnote{\label{foot:generalization}The proof is based on
  Eq.~(\ref{eq:TkkHo}), which can be seen to apply separately to the
  energy-momentum tensor of the scalar fields and of the gauge fields: since
  $\imath_{k}\imath_{k}F^{\Lambda}=0$, we find that
  \begin{equation}
    \imath_{k}F^{\Lambda}
    \stackrel{\mathcal{H}}{=}
    f^{\Lambda}\hat{k}+\hat{v}^{\Lambda}\,,
  \end{equation}
  where $\hat{k}=k_{\mu}dx^{\mu}$ and
  $\hat{v}^{\Lambda}=v^{\Lambda}{}_{\mu}dx{\mu}$ where $v^{\Lambda}$ is a
  spacelike vector normal to $k$. Then
  \begin{equation}
    k^{\mu}k^{\nu}T_{\mu\nu}
      \stackrel{\mathcal{H}}{\sim}
    g_{xy}\imath_{k}\phi^{x}\imath_{k}\phi^{y}
    -I_{\Lambda\Sigma}v^{\Lambda}\cdot v^{\Sigma}
    \stackrel{\mathcal{H}}{\sim}
    0\,.
  \end{equation}
  The second term is non-negative because $I_{\Lambda\Sigma}$ is
  definite-negative and the vectors $v^{\Lambda}$ are spacelike in
  mostly-minus signature. Since $g_{xy}$ is positive-definite, the condition
  Eq.~(\ref{eq:ikdphihorizon}) must be satisfied by each scalar $\phi^{x}$.
  Om the other hand, find $v^{\Lambda}=0$ and the conditions
  Eqs.~(\ref{eq:ikFonH0}) must be satisfied for all $\Lambda$, that is
  \begin{equation}
    \label{eq:ikFMonH0}
      \imath_{k}F^{M}
       \stackrel{\mathcal{H}}{=}
      0\,.
  \end{equation}
}
that, the event horizon of stationary, axisymmetric black holes satisfying
this ansatz (if any) is the Killing horizon of $k$, which implies the
``synchronization condition'' Eq.~(\ref{eq:synchronizationcondition})
Ref.~\cite{Herdeiro:2024pmv} which can only be satisfied in the case $K=L$,
$\phi^{1}=\phi^{2}$.

As for the gauge fields, following the same steps as in
Section~\ref{sec-electricmagneticandgeneralizedsymmetry}, defining the gauge
fields $A^{M}$, $F^{M}=dA^{M}$, we find that the generalized ansatz implies

\begin{equation}
\label{eq:ijivarphiF}  
    \imath_{t,\varphi}F^{M}
     =
     -DP_{t,\varphi}{}^{M}\,,
\end{equation}

\noindent
where we have defined the gauge-covariant derivatives

\begin{equation}
     DP_{t}{}^{M} \equiv dP_{t}{}^{M} -\omega T^{M}{}_{N}A^{N}\,,
     \hspace{1cm}
     DP_{\varphi}{}^{M} \equiv dP_{\varphi}{}^{M} -m S^{M}{}_{N}A^{N}\,,
\end{equation}

\noindent
invariant under

\begin{equation}
  \delta_{\chi}A^{M}
  =
  d\chi^{M}\,,
  \hspace{1cm}
  \delta_{\chi}P_{t}{}^{M} =\omega T^{M}{}_{N}\chi^{N}\,,
  \hspace{1cm}
  \delta_{\chi}P_{\varphi}{}^{M} = m S^{M}{}_{N}\chi\,.
\end{equation}

Notice that, in black-hole spacetimes, for
$k=\partial_{t}-\Omega_{H}\partial_{\varphi}$ and using
Eq.~(\ref{eq:ikFMonH0}) we find that

\begin{equation}
  \omega T^{M}{}_{N} -\Omega_{H} m S^{M}{}_{N}
  =
  0\,,
\end{equation}

\noindent
which implies $T=S$ and the synchronization condition. Then, in the
Einstein--Maxwell case the momentum maps satisfy the standard momentum map
equation (\ref{eq:momentummapequationsymplectic}) which leads to the
generalized zeroth law for the electrostatic and magnetostatic potentials
$\Phi^{M}=P_{k}{}^{M}$.

Let us now consider the definitions of the different charges in this setup. 

Electric and magnetic charges are still defined by Eq.~(\ref{eq:QMdef}) and
satisfy Gauss laws. Thus, they are doomed to vanish identically in boson
stars, but not necessarily in black-hole spacetimes.

Let us now consider the scalar charges. The standard symmetry arguments
leading to the expression Eq.~(\ref{eq:QAkcharges}) are valid and lead to the
same expressions except for the charges associated to the generators $K,L$ and
$T,S$ involved in the generalized symmetry ansatz. Then, except in these two
cases, those charges must vanish in boson stars and are subject to no-hair
theorems in asymptotically-flat black-hole spacetimes
\cite{Ballesteros:2023iqb}. The two exceptions fall in the general case
considered in Appendix~\ref{sec-generalisymmetricansatzscalarcharges} and,
since the global symmetries they are associated with commute, they have the
same form as in the standard case and must also vanish for boson stars and are
subject to no-hair theorems.

The charges $\mathbf{Q}_{U\, k}$ do not need to vanish, though, and should be
treated as in Section~\ref{sec-Estheoryscalarpotential}
\cite{Ballesteros:2023muf}.

Let us now consider the Komar charge. The Noether--Wald charge can be found as
in the symmetric case and takes the form \cite{Ballesteros:2023iqb}

\begin{equation}
  \mathbf{Q}[\xi]
  =
  \star (e^{a}\wedge e^{b})P_{\xi\, ab} +P_{\xi}{}^{\Lambda}F_{\Lambda}\,.
\end{equation}

By construction, it satisfies

\begin{equation}
  \label{eq:dQxionshell}
  d\mathbf{Q}[\xi]
  \doteq
  \mathbf{\Theta}(\varphi,\delta_{\xi}\varphi)
  +\imath_{\xi}\mathbf{L}\,,
\end{equation}

\noindent
where, in this theory

\begin{subequations}
  \begin{align}
    \mathbf{\Theta}(\varphi,\delta_{\xi}\varphi)
    & =
      -\star (e^{a}\wedge e^{b})\wedge \delta_{\xi} \omega_{ab}
      +g_{xy}\star d\phi^{x}\delta_{\xi}\phi^{y}
      +F_{\Lambda}\wedge \left(\imath_{\xi}F^{\Lambda}+dP_{\xi}{}^{\Lambda}\right)\,,
    \\
    & \nonumber \\
    \mathbf{L}
    & \doteq
      -\tfrac{1}{2}F^{\Lambda}\wedge F_{\Lambda}
      -\star V
      \,.
  \end{align}
\end{subequations}

For $\xi=k$, where $k$ is some Killing vector

\begin{subequations}
  \begin{align}
    \mathbf{\Theta}(\varphi,\delta_{k}\varphi)
    & =
      -\vartheta_{k}{}^{A}\left(K_{A\, x}\star d\phi^{x}
      -T_{A}{}^{\Lambda}{}_{M}F_{\Lambda}\wedge A^{M}\right)\,,
    \\
    & \nonumber \\
    \imath_{k}\mathbf{L}
    & \doteq
     \tfrac{1}{2}\left(dP_{k}{}^{\Lambda}\wedge F_{\Lambda}
      +dP_{k\, \Lambda} \wedge F^{\Lambda}\right)
      \nonumber \\
    & \nonumber \\
    & \hspace{.5cm}
      -\tfrac{1}{2}\vartheta_{k}{}^{A}
      \left(T_{A}{}^{\Lambda}{}_{M}F_{\Lambda}\wedge A^{M} 
     +T_{A\, \Lambda\, M}  F^{\Lambda}\wedge A^{M} \right)
      -\imath_{k}\star V\,,
  \end{align}
\end{subequations}

\noindent
so that

\begin{equation}
  \mathbf{\Theta}(\varphi,\delta_{k}\varphi)
  +\imath_{k}\mathbf{L}
  \doteq
  d\left\{  \tfrac{1}{2}\left(P_{k}{}^{\Lambda}\wedge F_{\Lambda}
    +P_{k\, \Lambda} \wedge F^{\Lambda}\right)\right\}
      +\vartheta_{k}{}^{A}J_{A}
      -\imath_{k}\star V\,,
\end{equation}

\noindent
where the on-shell-closed NGZ 3-form currents $J_{A}$ have the form given in
Eq.~(\ref{eq:JIdef}) for the subset of indices $A,B,\ldots$ corresponding to
the invariances of the scalar potential.
    
Substituting this expression in Eq.~(\ref{eq:dQxionshell}) we find the
identity

\begin{equation}
  d\left\{\star (e^{a}\wedge e^{b})P_{k\, ab}
    +\tfrac{1}{2}\left[P_{k}{}^{\Lambda}\wedge F_{\Lambda}
      -P_{k\, \Lambda} \wedge F^{\Lambda}\right]\right\}
  =
      \vartheta_{k}{}^{A}J_{A} -\imath_{k}\star V\,.  
\end{equation}

As usual, we can define on-shell the 2-forms $\mathcal{J}_{A}$ and
$\mathcal{V}_{k}$

\begin{subequations}
  \begin{align}
    J_{A}
    & \doteq
      d\mathcal{J}_{A}\,,
    \\
    & \nonumber \\
    \imath_{k}\star V
    & \doteq
      d\mathcal{V}_{k}\,,
  \end{align}
\end{subequations}

\noindent
whose form depends on the solution on which they are evaluated.  Notice that
the 2-forms $J_{A}$ do not satisfy a Gauss law and are different from the
2-form charges $\mathbf{Q}_{A}[k]$, which depend on the Killing vector and
satisfy a Gauss law.  The combinations $\vartheta_{k}{}^{A}J_{A}$ for
$k=\partial_{t},\partial_{\varphi}$ are just $\omega J_{K}$ and $mJ_{L}$ and
we will call the corresponding 2-forms $\omega \mathcal{J}_{K}$ and
$m\mathcal{J}_{L}$

The on-shell-closed generalized Komar charges is, therefore,

\begin{equation}
  \label{eq:generalizedKomarchargeEMstheorygeneralized}
  \mathbf{K}[k]
  =
  -\frac{1}{16\pi G_{N}{}^{(4)}}\left\{\star (e^{a}\wedge e^{b})P_{k\, ab}
  +\tfrac{1}{2}\left[P_{k}{}^{\Lambda}F_{\Lambda}
    -P_{k\, \Lambda}F^{\Lambda}\right]
  -\vartheta_{k}{}^{A}\mathcal{J}_{A}
  -\mathcal{V}_{k}\right\}\,.
\end{equation}

We can always derive a Smarr formula by integrating this generalized Komar
charge over the relevant boundaries. In boson-star spacetimes, though, it is
not clear how to identify generically the results in terms of conserved
charges and potentials. Since the electric and magnetic charges vanish, it is
likely that the terms containing the gauge field strengths will not contribute
to the integrals at infinity, but the other two terms can contribute,
explaining the existence of boson star solutions like those constructed in
Refs.~\cite{Herdeiro:2014goa,Herdeiro:2015gia,Cano:2023bpe} for instance.

In black-hole spacetimes, since, as we have explained, we must have $K=L$ and
$T=S$, we find that

\begin{equation}
  \vartheta_{k}{}^{A}\mathcal{J}_{A}
  =
  \vartheta_{t}{}^{A}\mathcal{J}_{A}
  -\Omega_{H}
  \vartheta_{\varphi}{}^{A}\mathcal{J}_{A}
  =
  \left(\omega -\Omega_{H}m\right) K
  =
  0\,,
\end{equation}

\noindent
by virtue of the synchronization condition
Eq.~(\ref{eq:synchronizationcondition}) and the next-to-last term in
Eq.~(\ref{eq:generalizedKomarchargeEMstheorygeneralized}) vanishes
identically. As we have discussed, the synchronization condition also ensures
that the momentum maps are constant over the horizon and we can proceed as in
the symmetric case, obtaining exactly the same Smarr formula
Eq.~(\ref{eq:Smarrformula3}). This formula allows for rotating black-hole
solutions as those constructed in
Refs.~\cite{Herdeiro:2014goa,Herdeiro:2015gia}.

\section{Discussion}
\label{sec-discussion}

In this paper we have shown how to construct 2-form charges satisfying Gauss
laws and how they can be used to restrict (or forbid) the existence of
boson-star or black-hole solutions. We have pad special attention to the
construction of the generalization of the standard Komar charge of General
Relativity from which Smarr formulas can be found. We have considered several
cases of increasing complexity leaving outside our scope theories of the
Proca--Higgs type which can be constructed by gauging the symmetries of the
scalars, eliminating some of them to obtain mass terms trough the
St\"uckleberg mechanism and also Yang-Mills fields. The former give rise to
``Proca--Higgs balls'', stars and black holes
\cite{Herdeiro:2023lze,Herdeiro:2024pmv} and the later to global monopoles
\cite{Harvey:1991jr,Gauntlett:1992nn,Chamseddine:1997nm,Huebscher:2007hj}. it
should be possible to extend the methods developed here to study these two
kinds of solutions, as well to extend them to asymptotically-AdS solutions
\cite{Anabalon:2022aig} for which one can use the positive energy theorem of
Ref.~\cite{Anabalon:2023oge}. Work in this direction is already under way
\cite{kn:BFMOP}.

We have also studied the generalized symmetric ansatz used to construct all
the known boson start solutions, considering a generic case with an arbitrary
group of isometries and global symmetries. We have argued that this ansatz
should be understood as a different implementation of the spacetime symmetries
on the matter fields and not as a breaking of those symmetries
(``non-inheritance'' \cite{Smolic:2015txa}).

The obvious similarity that we have found between the integrability condition
of this ansatz and the quadratic constraint of the embedding tensor formalism
is quite remarkable and calls for further study. It should be noticed that
this ansatz has been extensively used in the context of generalized
dimensional reduction and that it is also related to the construction of
``U-folds.'' These intriguing connections deserve further study since they may
lead to a complementary understanding of the reasons for the existence of the
boson star and hairy black hole solutions considered in the literature.

\section*{Acknowledgments}

This work has been supported in part by the MCI, AEI, FEDER (UE) grants
PID2021-125700NB-C21 (``Gravity, Supergravity and Superstrings'' (GRASS)), and
IFT Centro de Excelencia Severo Ochoa CEX2020-001007-S.  The work of RB has
also been supported by the National Agency for Research and Development [ANID]
Chile, Doctorado Nacional, under grant 2021-21211461 and by PUCV, Beca
Pasant\'{\i}a de Investigaci\'on.  TO wishes to thank M.M.~Fern\'andez for her
permanent support.

\appendix

\section{General considerations on the generalized symmetric ansatz}
\label{sec-generalconsiderations}

In this appendix we want to study the generalized symmetric ansatz for
arbitrary isometry and global groups. We will mainly focus on scalar fields,
for the sake of simplicity. Thus, we assume

\begin{enumerate}
\item The existence of a group of isometries of the spacetime metric generated
  by the Killing vector fields $k_{m}\equiv k_{m}{}^{\mu}(x)\partial_{\mu}$,
  with Lie brackets

  \begin{equation}
    \label{eq:spacetimeLiealgebra}
    [k_{m},k_{n}] = f_{mn}{}^{p}k_{p}\,,
  \end{equation}

\noindent
which may simply be a subgroup of the complete isometry group. The isometries
we are considering may be those of an internal space, if we are interested in
a dimensional compactification ansatz, or just part of the ansatz for a
boson-star, black-hole or any other kind of solution.

The diffeomorphisms generated by those Killing vector fields act on the matter
fields, here represented by scalar fields $\phi^{x}$ parametrizing some target
space, as

\begin{equation}
  \delta_{m}\phi^{x}
  =
  -\pounds_{k_{m}}\phi^{x}\,,
\end{equation}

\noindent
where $\pounds_{k_{m}}$ is the Lie derivative with respect to the vector field
$k_{m}$

\begin{equation}
  \pounds_{k_{m}}\phi^{x}
  =
  \imath_{k_{m}}\phi^{x}
  \equiv
  \imath_{m}\phi^{x}\,.
\end{equation}

\item The existence of a global symmetry group acting on the matter fields. On
  the scalar fields that we are considering as an example, the generators are 

\begin{equation}
  \delta_{I}\phi^{x}
  =
  K_{I}^{x}(\phi)\,,
\end{equation}

\noindent
where the $K_{I}^{x}(\phi)$ are Killing vectors of the target space metric
$g_{xy}(\phi)$ satisfying the Lie algebra

\begin{equation}
  \label{eq:targetspaceLiealgebra}
  [K_{I},K_{J}]
  =
  f_{IJ}{}^{K}K_{K}\,.
\end{equation}

\end{enumerate}

The generalized symmetric ansatz assumes that the scalar fields are not
symmetric under the infinitesimal general coordinate transformations (GCTs)
generated by the Killing vectors of the spacetime metric, in the naive sense

  \begin{equation}
    \pounds_{m}\phi^{x}
    \equiv
    \pounds_{k_{m}}\phi^{x}
    =
    0\,,
  \end{equation}

\noindent
but in a generalized sense \cite{Yazadjiev:2024rql}
  
\begin{equation}
    \label{eq:generalizedansatz}
    \delta_{m}\phi^{x}
    \equiv
    -\pounds_{m}\phi^{x}
    +\vartheta_{m}{}^{I}K_{I}{}^{x}
    =
    0\,.
\end{equation}

Here the $\vartheta_{m}{}^{I}$ are constants (we call them \textit{shift
  constants}) and the above equation indicates that the scalars are invariant
under the GCTs generated by the spacetime Killing vectors up to a global
symmetry generated by a certain combination of the target-space Killing
vectors $K_{I}$.

\subsection{Consistency condition}
\label{sec-consistencycondition}

The ansatz Eq.~(\ref{eq:generalizedansatz}) has to satisfy a consistency (or
integrability) condition: if we act with the Lie derivative with respect to a
different spacetime Killing vector, we find

\begin{equation}
  \label{eq:consistency1}
  \pounds_{m}\pounds_{n}\phi^{x}
  =
  \vartheta_{n}{}^{I}\pounds_{m}K_{I}{}^{x}
  =
  \vartheta_{n}{}^{I}\partial_{y}K_{I}{}^{x}\pounds_{m}\phi^{y}
  =
  \vartheta_{n}{}^{I}\vartheta_{m}{}^{J}K_{J}{}^{y}\partial_{y}K_{I}{}^{x}\,,
\end{equation}

\noindent
and, antisymmetrizing in $m,n$ and using the definition of the Lie bracket, we
get, up to a factor of $1/2$,

\begin{equation}
  \label{eq:consistency2}
  [\pounds_{m},\pounds_{n}]\phi^{x}
  =
  -
  \vartheta_{m}{}^{I}\vartheta_{n}{}^{J}[K_{I},K_{J}]^{x}
  =
  -
  \vartheta_{m}{}^{I}\vartheta_{n}{}^{J}f_{IJ}{}^{K}K_{K}{}^{x}\,,
\end{equation}

\noindent
where we have used Eq.~(\ref{eq:targetspaceLiealgebra}).  Using the
fundamental property of the Lie derivative, Eq.~(\ref{eq:spacetimeLiealgebra})
and the linearity of the Lie derivative, we find that, 

\begin{equation}
    \label{eq:consistency3}
  [\pounds_{k_{m}},\pounds_{k_{n}}]
  =
  \pounds_{[k_{m},k_{n}]}
  =
  \pounds_{f_{mn}{}^{p}k_{p}}
  =
  f_{mn}{}^{p}\pounds_{k_{p}}\,,
\end{equation}

\noindent
and plugging this relation into Eq.~(\ref{eq:consistency2}) and using the
ansatz again, we arrive to the following relation between structure constants
and shift constants

\begin{equation}
  \label{eq:consistencycondition}
  f_{mn}{}^{p}\vartheta_{p}{}^{K}
  =
  -\vartheta_{m}{}^{I}\vartheta_{n}{}^{J}f_{IJ}{}^{K}\,.
\end{equation}

We notice that this relation is formally identical to the so-called quadratic
constraint of the embedding tensor formalism.\footnote{See,
  \textit{e.g.}~Eq.~(3.57) in Ref.~\cite{Trigiante:2016mnt}.}  In that
formalism, the consistency condition Eq.~(\ref{eq:consistencycondition}) can
be seen to arise from the requirement that the embedding tensor be
invariant. Here, if we view the shift constants $\vartheta_{m}{}^{I}$ as
objects an adjoint index $m$ and another index $I$ in some other
representation r of the spacetime symmetry subgroup, and transforming with
matrices

\begin{equation}
  \Gamma_{\rm Adj}(T_{m})^{p}{}_{n} \equiv f_{mn}{}^{p}\,,
  \hspace{1cm}
  \Gamma_{\rm r}(T_{m})^{K}{}_{J} \equiv \vartheta_{m}{}^{I}f_{IJ}{}^{K}\,,  
\end{equation}

\noindent
invariance means

\begin{equation}
  \delta_{m}\vartheta_{n}{}^{I}
  =
  \Gamma_{\rm Adj}(T_{m})^{p}{}_{n}\vartheta_{p}{}^{I}
  +\Gamma_{\rm r}(T_{m})^{I}{}_{J} \vartheta_{n}{}^{J}
  =
  f_{mn}{}^{p}\vartheta_{p}{}^{I}+ \vartheta_{m}{}^{J}\vartheta_{n}{}^{K}f_{JK}{}^{I}
  =
  0\,,
\end{equation}

\noindent
which is the consistency condition Eq.~(\ref{eq:consistencycondition}).

This consistency condition imposes strong constraints on the possible
ansatzs. Let us consider, for example, a 4-dimensional spherically-symmetric
spacetime and let us focus on the generators of the SO$(3)$ isometry subgroup
with structure constants $f_{mn}{}^{p} = \varepsilon_{mnp}$,
$m,\ldots=1,2,3$. It is evident that the shift constants can only be
non-trivial if $f_{IJ}{}^{K}\neq 0$ and, therefore, one uses for the
generalized symmetric ansatz a non-Abelian subgroup of the target space
isometry group. In particular, for the very often considered case of a
massive, complex, Klein--Gordon scalar, which only has one available isometry,
all the components of the shift constants must vanish identically in the
spherically symmetric case and none of the generators of SO$(3)$ can be used
to define a generalized symmetric ansatz. However, if the spacetime is
stationary and axisymmetric, those two commuting spacetime symmetries can be
combined with the phase shifts of the Klein--Gordon
field as in Ref.~\cite{Herdeiro:2015gia}. This example was studied in
Ref.~\cite{Yazadjiev:2024rql}

Observe that a trivial way to satisfy the constraint is to use identical
spacetime and target-space Lie algebras and shift constants which are
proportional to Kronecker deltas.

\subsection{Scalar charges and the generalized symmetric ansatz}
\label{sec-generalisymmetricansatzscalarcharges}

As we have explained in the main text, in the standard symmetric case, one can
construct 2-form scalar charges satisfying a Gauss law using the invariance of
all the fields under the diffeomorphisms generated by Killing vector, which
implies the invariance of the on-shell-closed Noether--Gaillard--Zumino (NGZ)
3-form currents (see Eq.~(\ref{eq:constructionofscalarcharges}))
$\delta_{k}J_{I}$.\footnote{Here, $J_{I}$ is the NGZ 3-form current associated
  to the global generator labeled by $I$.} This leads to the on-shell
closedness of the interior products of the Killing vector $k_{m}$ and the NGZ
3-forms $d\imath_{m}J_{I}\doteq 0$, which we can use as 2-form scalar charges
$\mathbf{Q}_{m\, I}\equiv \imath_{m}J_{I}$.

This mechanism does not work in the generalized case, in which we have

\begin{equation}
  \begin{aligned}
  \delta_{m}J_{I}
  & =
    -d\imath_{m}J_{I}
    -\imath_{m} dJ_{I}
    +\vartheta_{m}{}^{J}\delta_{J}J_{I}
    \\
    & \\
    & \doteq
    -d\imath_{m}J_{I}
      +\vartheta_{m}{}^{J}f_{JI}{}^{K}J_{K}
      =
      0\,,
  \end{aligned}
\end{equation}

\noindent
except when one uses Abelian global symmetry groups.

In the non-Abelian case, though, we can define on-shell 2-forms $B_{I}$

\begin{equation}
  J_{I}
  \dot{\equiv}
  dB_{I}\,,
\end{equation}

\noindent
that can be seen as the duals of the scalar fields
\cite{Bandos:2016smv,Fernandez-Melgarejo:2023kwk} and we can rewrite the above
expression as a total derivative

\begin{equation}
    d\left(\imath_{m}dB_{I}
      -\vartheta_{m}{}^{J}f_{JI}{}^{K}B_{K}\right)
      \doteq
      0\,,
\end{equation}

\noindent
which allows us to define the on-shell-closed 2-form scalar charges

\begin{equation}
  \mathbf{Q}_{m\, I}
  \equiv
  \imath_{m}dB_{I} -\vartheta_{m}{}^{J}f_{JI}{}^{K}B_{K}\,,
  \hspace{1cm}
  d \mathbf{Q}_{m\, I}
  \doteq
  0\,.
\end{equation}

\section{The (restricted)  generalized zeroth law for the electrostatic and
  magnetostatic potentials}
\label{sec-generalizedzerothlaw}

If we define the electrostatic and magnetostatic potentials through the
momentum map equations (\ref{eq:momentummapequation}) and
(\ref{eq:dualmomentummapequation}) that we reproduce here for the sake of
convenience

\begin{subequations}
  \begin{align}
    \imath_{k}F+d\Phi
    & =
      0\,,\\
    & \nonumber \\
    \imath_{k}\tilde{F}+d\tilde{\Phi}
    & =
      0\,,
  \end{align}
\end{subequations}

\noindent
using the Killing vector $k=\partial_{t}-\Omega_{H}\partial_{\varphi}$ that
satisfies $k^{2}\stackrel{\mathcal{H}}{=}0$ and $k\stackrel{\mathcal{BH}}{=}0$,
we immediately find that these potentials are constant on $\mathcal{BH}$:

\begin{subequations}
  \begin{align}
    d\Phi
    & \stackrel{\mathcal{BH}}{=}
      0\,,
      \\
    & \nonumber \\
    d\tilde{\Phi}
    & \stackrel{\mathcal{BH}}{=}
      0\,.
  \end{align}
\end{subequations}
    
In Ref.~\cite{Elgood:2020svt} this property has been called the
\textit{restricted generalized zeroth law}, since it is a restriction of the
standard generalized zeroth law which says that they are constant over the
whole $\mathcal{H}$ to $\mathcal{BH}$. The generalized zeroth law follows from 
the restricted one and from the equations

\begin{subequations}
  \label{eq:ikdPk0}
  \begin{align}
  \imath_{k}d\Phi
    & =
      0\,,
      \\
    & \nonumber \\
     \imath_{k}d\tilde{\Phi}
    & =
      0\,,
  \end{align}
\end{subequations}

\noindent
which are obtained by taking the interior product of the momentum maps
equations with $k$.

The standard derivation of the generalized zeroth law\footnote{See, for
  instance~Ref.~\cite{Frolov:1998wf}.}  that does not relie on the existence
of a bifurcation surface but needs, instead, the Einstein equations from which
Eq.~(\ref{eq:TkkHo}) follows. Applying this identity to the energy-momentum
tensor of the Maxwell field we find\footnote{The Maxwell energy-momentum
  tensor is invariant under the replacement of $F$ by $\star F=\tilde{F}$.}

\begin{subequations}
  \label{eq:Temkk0onH}
  \begin{align}
  \imath_{k}F\cdot\imath_{k}F
  & \stackrel{\mathcal{H}}{=}
    0\,,
    \\
& \nonumber \\
  \imath_{k}\tilde{F}\cdot\imath_{k}\tilde{F}
  & \stackrel{\mathcal{H}}{=}
  0\,,
  \end{align}
\end{subequations}
which implies that $\imath_{k}F$ and $\imath_{k}\tilde{F}$ are null in
$\mathcal{H}$. Since $\imath_{k}\imath_{k}F=\imath_{k}\imath_{k}\tilde{F}=0$,
the only possibility is that

\begin{subequations}
  \label{eq:ikFpropotok}
  \begin{align}
  \imath_{k}F
  & \stackrel{\mathcal{H}}{\propto}
    \hat{k}\,,
    \\
    & \nonumber \\
  \imath_{k}\tilde{F}
  & \stackrel{\mathcal{H}}{\propto}
  \hat{k}\,,
  \end{align}
\end{subequations}

\noindent
where $\hat{k}$is the 1-form dual to the Killing vector $k$, that is
$\hat{k}\equiv k_{\mu}dx^{\mu}$. Using the momentum-map equations, it follows
that

\begin{subequations}
  \begin{align}
  \hat{k}\wedge d\Phi
    & \stackrel{\mathcal{H}}{=}
      0\,,
      \\
    & \nonumber \\
     \hat{k}\wedge d\tilde{\Phi}
    & \stackrel{\mathcal{H}}{=}
      0\,,
  \end{align}
\end{subequations}

\noindent
which, together with Eqs.~(\ref{eq:ikdPk0}) implies that $\Phi$ and
$\tilde{\Phi}$ are constant over $\mathcal{H}$.

Here it is important that these functions are defined as momentum maps with
respect to $k=\partial_{t}-\Omega_{H}\partial_{\varphi}$. The purely
electrostatic and magnetostatic potentials that would have been obtained in
$k=\partial_{t}$ do not have this property. The electric and magnetic charges
are the same on the horizon and at infinity because they satisfy Gauss laws
and they no longer need to vanish.

Finally, observe that, as a consequence of the generalized zeroth law,

\begin{subequations}
  \label{eq:ikFonH0}
  \begin{align}
    \imath_{k}F
    & \stackrel{\mathcal{H}}{=}
      0\,,
      \\
    & \nonumber \\
    \imath_{k}\tilde{F}
    & \stackrel{\mathcal{H}}{=}
      0\,.
  \end{align}
\end{subequations}

A generalization of these results when there are several gauge and scalar
fields can be found in footnote~\ref{foot:generalization} in
page~\pageref{foot:generalization}.

\section{A generalized stationary solution of the Maxwell equations in
  Minkowski spacetime}
\label{sec-trial}

In this appendix we want to show how to find solutions of the Maxwell
equations and Bianchi identities satisfying the generalized symmetric ansatz
Eqs.~(\ref{eq:GSAinEM}) with $k=\partial_{t}$. It is convenient to work with
the gauge fields $A,\tilde{A}$ because the Maxwell equations and Bianchi
identities are automatically satisfied and we only need to demand the
self-duality condition Eq.~(\ref{eq:dualfieldstrengthdef}). Thus, we are going
to use the formulation of the generalized symmetric ansatz for gauge fields
Eqs.~(\ref{eq:GSAinEMforgaugefields}). We also make the ansatz/gauge choice
$\chi_{k}=\tilde{\chi}_{k}=0$ that reduces
Eqs.~(\ref{eq:GSAinEMforgaugefields}) to

\begin{equation}
  \partial_{t}
  \left(
    \begin{array}{c}
      A \\ \tilde{A} \\
    \end{array}
  \right)
=
  \left(
    \begin{array}{c}
    \omega  \tilde{A} \\ -\omega A \\
    \end{array}
  \right)\,,
\end{equation}

\noindent
which, working in Cartesian coordinates, can be solved by

\begin{equation}
  A_{t}
  =
  \tilde{A}_{t}
  =
  0\,,
  \hspace{1cm}
  A_{m}
  =
  \cos{(\omega t)}\, f_{m}(x)\,,
  \hspace{1cm}
  \tilde{A}_{m}
  =
  \sin{(\omega t)}\, f_{m}(x)\,,
  \,\,\,\,
  m,n,p,\ldots=1,2,3\,.
\end{equation}

\noindent
The solution will satisfy the self-duality condition
Eq.~(\ref{eq:dualfieldstrengthdef}) if the time-independent $f_{m}(x)$
satisfies the equation

\begin{equation}
  \varepsilon_{mnp}\partial_{n}f_{p}
  =
  -\omega f_{m}\,,
  \,\,\,\,
  \Rightarrow
  \,\,\,\,
  \partial_{m}f_{m}
  =
  0\,.
\end{equation}

This equation is solved in spherical coordinates by

\begin{equation}
  f_{r}
  =
  0\,,
  \hspace{1cm}
  f_{\theta}
  =
  \frac{A \cos{(\omega r)} +B \sin{(\omega t)}}{\sin{\theta}}\,,
  \hspace{1cm}
  f_{\varphi}
  =
  -A \sin{(\omega r)} +B \cos{(\omega r)}\,,
\end{equation}

\noindent
where $A$ and $B$ are integration constants.

This solution oscillates in time and space and the components of the gauge
fields are sines or cosines of $\omega(t\pm r)$. The components
$F_{\theta\varphi}$ and $\tilde{F}_{\theta\varphi}$ vanish identically and so
do the electric and magnetic charges. The non-vanishing components of the
energy-momentum tensor do not decay fast enough at infinity but they are
time-independent, as expected:

\begin{equation}
  T_{tt}=T_{rr}
  \sim
  -\frac{\omega^{2}(A^{2}+B^{2})}{r^{2}\sin^{2}{\theta}}\,,
  \hspace{1cm}
  T_{\varphi\varphi} = -\sin^{2}{\theta} T_{\theta\theta}
  \sim
  4\omega^{2}AB \sin{(\omega r)} \cos{(\omega r)}\,.
\end{equation}


\end{document}